\documentstyle[12pt,epsfig]{article}
\def\thebibliography#1{\leftline{\Large\it References}\list
  {[\arabic{enumi}]}{\settowidth\labelwidth{[#1]}\leftmargin\labelwidth
    \advance\leftmargin\labelsep
    \usecounter{enumi}}
    \def\newblock{\hskip .11em plus .33em minus .07em}
    \sloppy\clubpenalty4000\widowpenalty4000}

\newcommand{\be}{\begin{eqnarray}}
\newcommand{\ba}{\begin{array}}
\newcommand{\ea}{\end{array}}
\newcommand{\ee}{\end{eqnarray}}

\newcommand{\dslash}{\partial \hskip -0.5em /}
\newcommand{\kslash}{k \hskip -0.5em /}
\newcommand{\Dslash}{D \hskip -0.7em /}
\newcommand{\tr}{{\rm tr}}
\newcommand{\Tr}{{\rm Tr}}

\newcommand{\La}{{\cal L}}
\newcommand{\A}{{\cal A}}

\newcommand{\ie}{{\it i.e.}\ }
\newcommand{\eg}{{\it e.g.}\ }

\newcommand{\smalllineskip}{\baselineskip=12pt}

\begin{document}
\rightline{\today}
\rightline{UNITU-THEP-2/1999}
\rightline{MIT--CTP--2807}
\rightline{hep-ph/9902322}
\vskip 0.8truecm
\centerline{\Large\bf Nucleon Structure Functions in the}
\vskip0.4cm
\centerline{\Large\bf
Three Flavor NJL Soliton Model$^{\textstyle*}$}
\baselineskip=16 true pt
\vskip 1.0cm
\centerline{O.\ Schr\"oder$^{\rm a}$, H. Reinhardt$^{\rm a}$ 
and H.\ Weigel$^{{\textstyle\dagger}\,{\rm a,b}}$}
\vskip 0.5cm
\begin{center}
$^{\rm a}$Institute for Theoretical Physics,
T\"ubingen University\\
Auf der Morgenstelle 14,
D-72076 T\"ubingen, Germany\\
\vskip0.5cm
$^{\rm b}$Center for Theoretical Physics\\
Laboratory of Nuclear Science and Department of Physics\\
Massachusetts Institute of Technology\\
Cambridge, Ma 02139
\end{center}
\vskip 0.8cm
\baselineskip=16pt
\centerline{\bf ABSTRACT}
\vskip 0.4cm
\centerline{\parbox[t]{15.2cm}{\baselineskip=20pt
We study the relevance of strange degrees of freedom for nucleon
structure functions. For this purpose we employ the three 
flavor generalization of the collective quantization approach to 
the chiral soliton of the bosonized Nambu--Jona--Lasinio model. 
Contrary to many other soliton models the hadronic tensor is tractable in 
this model. By applying the Bjorken limit to the hadronic tensor we 
extract the leading twist contributions to the nucleon structure 
functions at the low energy scale at which the model is assumed to 
approximate QCD. After transforming to the infinite momentum frame 
and performing the DGLAP evolution program to these structure functions 
we compare with available data for deep inelastic electron--nucleon 
scattering.  }}
\vskip 0.5cm
\leftline{\it PACS: 11.30.Cp, 12.39.Ki.}

\vfill
\noindent
--------------

\noindent
$^{\textstyle*}$\hskip-0.03cm
This work is supported in parts by funds provided by the 
U.S. Department of Energy (D.O.E.) under cooperative research
agreement \#DF--FC02--94ER40818 and by the Deutsche 
Forschungsgemeinschaft (DFG) under contract We 1254/3-1 and
the {\it Graduiertenkolleg Hadronen und Kerne} at 
T\"ubingen University.\\
$^{\textstyle\dagger}$\hskip-0.02cm
{Heisenberg--Fellow}
\eject

\baselineskip=16pt

\leftline{\Large\it 1. Introduction}
\medskip

In this paper we present the computation of the leading twist pieces 
of both the unpolarized and polarized nucleon structure functions in 
the three flavor Nambu--Jona--Lasino (NJL) \cite{Na61} chiral soliton 
model \cite{Re88,Al96,We92}. Applying functional bosonization techniques 
\cite{Eb86} this model for the quark flavor dynamics can be 
straightforwardly expressed in terms of meson degrees of freedom. The 
identification of baryons as solitons of effective meson theories is 
motivated by generalizing quantum chromodynamics (QCD) to a large 
number ($N_C)$ of color degrees of freedom \cite{Wi79}. Essentially this 
generalization shows that the baryon properties exhibit the same 
dependence on $N_C$ as expected for soliton solutions of the effective 
meson theory. In addition to this analogy there are various 
reasons based on phenomenological observations to consider soliton models 
suitable for studying baryon properties. For example, such models quite 
successfully describe many aspects of static baryon properties 
\cite{Sk61,Ho84,Me88,We96r}. This is in particular the case for observables 
related to the {\it proton spin puzzle} \cite{Br88} which is directly 
related to the polarized structure functions of the nucleon. Working in 
the three flavor extension of the model provides us with the advantageous 
feature of direct access to the strange quark contributions to structure 
functions even though the {\it net} strangeness of the nucleon vanishes. 
Hence we can gauge the importance of these degrees of freedom for 
${\underline{\rm dynamical}}$ properties of the nucleon. This is 
particularly interesting because the analyzes of experiments performed at 
CERN and SLAC have indicated that the spin of the strange quarks inside 
the nucleon yields a sizable contribution to the total nucleon spin.

Unfortunately there are a number of soliton models available. The reason 
for adopting the NJL chiral soliton model lies in the fact that the main 
ingredient in any structure function calculation is a bilocal commutator 
of hadronic currents as can be observed from the defining equation 
(\ref{hten1}) of the hadronic tensor. It appears that its calculation 
in models with elementary meson fields is infeasible. However, it is 
much less involved in a model defined in terms of quark fields because 
the Compton amplitude, whose absorptive part equals this commutator, can 
be traced in the process of functional bosonization. 
Formally this process provides a unique regularization of the 
quark loop. Leaving aside the cumbersome issue of regularization 
in the NJL model, the leading twist contribution to the absorptive 
part of the Compton amplitude becomes as simple as a bilocal and bilinear 
operator of the quark fields.

In the present approach to the three flavor NJL--model we do 
not include the 't Hooft determinant \cite{tHo76} which, when 
included \cite{Re88a}, provides a mass for the 
singlet component of the $\eta$--meson and parameterizes the mixing 
of the uncharged pseudoscalar mesons. The reason for ignoring this 
piece of the action is that pseudoscalar flavor singlets couple 
only weakly to the chiral soliton, if at all \cite{Br88}. 
In the soliton approach the influence of the 't Hooft determinant
on nucleon properties is hence negligible. It should however
be remarked that at a sub--leading order in $1/N_C$ this term might 
give some contributions of the strange quarks to nucleon 
properties due to a mixing of non--strange and strange 
constituent quarks \cite{Be87}.

We should mention that there is an exhaustive number of model 
calculations for nucleon structure functions. The models applied
range from various kinds of bag models \cite{Ja75,Be87a}
over cloudy bag models \cite{Sa88} and meson cloud models
\cite{Sz96} to (diquark) spectator models \cite{Me91}, just
to name a few. In the context of structure functions those model 
calculations may eventually be somewhat more reliable than soliton 
model results. After all, many of those models have particularly
been set up to study structure functions. On the other hand 
soliton models are very attractive not only because of the above 
mentioned connection to QCD in the limit of large $N_C$ and the
possibility to generalize them to three flavors but over 
and above they provide a very comprehensive picture of baryons ranging 
way above static properties of nucleons from meson--nucleon scattering 
\cite{Wa84,Me88} over systems with a heavy quark \cite{Ca85} to 
applications in the context of nuclear matter \cite{Ja88}. Thereby 
these models use only very a limited number of parameters. The present 
study apparently serves to further complete this comprehensive picture.

This paper is organized as follows: In section 2 we briefly 
review the NJL chiral soliton model and discuss the inclusion of 
strange degrees of freedom within the collective approach. Effects 
due to flavor symmetry breaking are incorporated into the collective 
nucleon wave--function by a generalization of the Yabu--Ando method 
\cite{Ya88,We96r}. We then explain the calculation of the hadronic 
tensor for localized field configurations \cite{Ja75}. Employing 
pertinent projection techniques the structure functions are extracted 
from this tensor. In this section we furthermore include a brief 
discussion on the relevance of boosting to the infinite momentum frame
\cite{Ja80,Ga97}. In the $1/N_C$ expansion scheme the dynamical response 
of the soliton to the momentum transfer is omitted which results in a 
violation of Lorentz covariance. Fortunately these omissions become 
irrelevant in the infinite momentum frame. We also discuss the valence 
quark approximation \cite{We96} to the structure function calculation 
which is to be considered as an approximation to avoid ambiguities 
inevitably arising when regularizing the vacuum part to 
bilocal--bilinear quark operators. Some of the previous approaches 
\cite{Di96,Wa98} to the two flavor structure functions adopted a 
once--subtracted Pauli--Villars type scheme. This, however, is not 
sufficient to render finite the quadratic divergences contained in the 
model. As a consistent incorporation of flavor symmetry breaking in the 
three flavor model demands the solution of the quadratically divergent 
gap--equation a single subtraction will not be suited for present 
study\footnote{In the case of the structure functions the absorptive 
part of the Compton amplitude -- a matrix element of an operator which 
is quartic in the quark fields -- appears to be the suitable starting 
point for a consistent regularization with two subtractions.}.
In section 3 we present our results for the unpolarized 
structure functions for both the electric and the strange current. 
Subsequently we discuss the polarized structure function $g_1$ in section 4. 
In particular we reflect on the electric and strange pieces in the light of 
the {\it proton spin puzzle}. In section 5 we will perform the DGLAP program 
to evolve the predicted structure functions from the low energy scale at 
which the model supposedly approximates QCD to the energy scale encountered 
in the respective experiments. In section 6 the numerical results will 
be presented and compared to data from CERN and SLAC experiments. Finally 
section 7 serves to summarize and conclude our study. Technicalities of 
the calculations are relegated to appendices.

Some of the results presented in this paper on the strangeness 
contribution to the polarized structure function $g_1(x)$ 
have already been published earlier \cite{Sch98}.

\bigskip
\leftline{\Large\it 2. \parbox[t]{14cm}
{The Hadronic Tensor from The Chiral Soliton
in the Three Flavor NJL Model}}
~
\medskip

We split up this section into two pieces. In the first one we will
review the treatment of the chiral soliton in the three flavor NJL model
while the second one contains the discussion of the hadronic 
tensor $W_{\mu\nu}$ in this model.

\medskip

\leftline{\large\it 2a. The Chiral Soliton in the
Three Flavor NJL Model}

\medskip
The basis of our considerations is the NJL model Lagrangian \cite{Na61}
\be
\La = \bar q (i\dslash - \hat m^0 ) q +
      2G_{\rm NJL} \sum _{i=0}^{N_f^2-1}
\left( (\bar q \frac {\lambda ^i}{2} q )^2
      +(\bar q \frac {\lambda ^i}{2} i\gamma _5 q )^2 \right) 
\label{NJL}
\ee
which is chirally symmetric for a vanishing current quark mass 
matrix $\hat m^0$. In eq (\ref{NJL}) $q$ denotes the quark field.
The matrices $\lambda ^i/2$ are the generators of the flavor
group $U(N_f)$. We will consider the case of three flavors ($N_f=3$) 
and neglect isospin breaking, \ie the current quark mass matrix acquires
the form $\hat m^0={\rm diag}(m^0,m^0,m^0_s)$ as $m^0=m_u^0=m_d^0$  is 
assumed. The NJL coupling constant $G_{\rm NJL}$ will later be
determined from meson properties.

Applying functional integral bosonization techniques the model
(\ref{NJL}) can be rewritten in terms of composite meson fields 
\cite{Eb86}.  The corresponding effective action is given by the sum 
$\A_{NJL}=\A_F+\A_m$ of a fermion determinant
\be
\A_F=\Tr\log(i\Dslash)=\Tr\log\left(i\dslash-(P_RM+P_LM^{\dag})\right)
\label{fdet}
\ee
and a purely mesonic part
\be
\A_m=\int d^4x\left(-\frac{1}{4G_{\rm NJL}}
\tr(M^{\dag}M-\hat m^0(M+M^{\dag})+(\hat m^0)^2)\right).
\label{ames}
\ee
Here $P_{R,L}=(1\pm \gamma _5)/2$ denote the projectors onto right-- 
and left--handed quark fields, respectively. The complex matrix 
$M=S+iP$ parameterizes the scalar and pseudoscalar meson fields.

Obviously, the action $\A_F$ is quadratically divergent and needs to be
regularized. We will apply the $O(4)$ invariant proper time regularization 
to $\A_F$ in Euclidean space which then becomes a complex quantity 
$\A_F=\A_R+\A_I$. Proper time regularization amounts to replacing 
the real part of the fermion determinant $\A_R$ by a parameter integral
\be
\A_R=\frac{1}{2}\Tr\log\left(\Dslash_E^{\dag}\Dslash_E\right)
\longrightarrow -\frac{1}{2}\int_{1/\Lambda ^2}^\infty
\frac{ds}s\Tr\exp\left(-s\Dslash_E^{\dag}\Dslash_E\right) 
\label{arreg}
\ee
which becomes exact in the limit $\Lambda\to \infty$ up to 
an irrelevant additive constant. The imaginary part
\be
\A_I=\frac{1}{2}\Tr\log\left((\Dslash_E^{\dag})^{-1}\Dslash_E\right)
\label{af3}
\ee
is finite and does not undergo regularization in order not to 
spoil the anomaly.

To study the three flavor extension of the model it is suitable
to introduce the following parameterization for the meson fields
\be
M=\xi\Sigma\xi\, .
\label{defm}
\ee
The matrix $\Sigma$ is Hermitian whereas $\xi$ is unitary. In this 
parameterization the chiral ${\rm SU}_L(3)\times{\rm SU}_R(3)$
transformation $M\to L M R^\dagger$ with constant matrices 
$L$ and $R$ is realized via $\Sigma\to K\Sigma K^\dagger$
and $\xi\to L\xi K^\dagger = K\xi R^\dagger$. The latter
equation determines the unitary matrix $K$ \cite{Ca69}.

Varying the action with respect to $\Sigma$ yields the Schwinger--Dyson 
or gap equation which determines the vacuum expectation value (VEV), 
$\langle \Sigma \rangle ={\rm diag}(m,m,m_s)$ and relates it to the 
quark condensate $\langle \bar q q \rangle _i$
\be
m_i & = & m_i^0+m_i^3\frac{N_C G_{\rm NJL}}{2\pi^2}
\Gamma\left(-1,(\frac{m_i}{\Lambda})^2\right)
=m_i^0-2G_{\rm NJL} \Big\langle \bar q q \Big\rangle _i .
\label{conmass}
\ee
Here $m_i=m,m_s$ denote the constituent masses of non--strange 
and strange quarks, respectively. The appearance of an incomplete
Gamma--function of the order $-1$ reflects the fact that the 
model is quadratically divergent in the ultraviolet. 
 
Space--time dependent fluctuating pseudoscalar meson
fields $\eta_a(x)$ are introduced via
\be
\xi_f(x)={\rm exp}\left(i\sum_{a=1}^8\eta_a(x)\lambda_a/2\right) .
\label{defeta}
\ee
Expanding the effective action up to second order in the fluctuations
$\eta_a(x)$ allows us (after rotating back to Minkowski space) to extract 
the inverse propagator for the pseudoscalar mesons 
$P_{ij}=\sum_{a=1}^{N_f^2-1}\eta^a\lambda^a_{ij}$ \cite{We92}
\be
D_{ij,kl}^{-1}(q^2)=
\Big(-\frac{(m_i^0+m_j^0)(m_i+m_j)}{2G_{\rm NJL}}-\Pi_{ij}(q^2)\Big)
\delta_{il}\delta_{kj}.
\label{prop}
\ee
The polarization operator $\Pi_{ij}(q^2)$ is given by
\be
\Pi_{ij}(q^2)=-2q^2f_{ij}^2(q^2)+2(m_i-m_j)^2f_{ij}^2(q^2)
-\frac{1}{2}(m_i^2-m_j^2)\Big(\frac{\langle\bar q q \rangle_i}{m_i}-
\frac{\langle\bar q q \rangle_j}{m_j}\Big)
\label{polar}
\ee
wherein
\be
f^2_{ij}(q^2)=\frac{1}{4}(m_i+m_j)^2\frac{N_c}{4\pi^2}\int_0^1dx\
\Gamma\Big(0,[(1-x)m_i^2+xm_j^2-x(1-x)q^2]/\Lambda^2\Big).
\label{fq}
\ee
The Bethe--Salpeter equation $D^{-1}(q^2)P=0$ which determines the 
physical meson masses $m_{\rm phys}$ is equivalent to the condition 
that the meson propagator acquires a pole:
\be
{\rm det}\left(D_{ij,kl}^{-1} (q^2=m_{\rm phys}^2)\right)=0\, .
\label{BSeq}
\ee
Note that $f^2_{ij}(q^2=m_{\rm phys}^2)$ is the corresponding 
on--shell meson decay constant. From eq (\ref{fq}) we can read 
off the pion decay constant 
\be
f_\pi ^2 =m^2 \frac {N_c}{4\pi ^2} \int _0^1 dx
\Gamma\Big(0,[m^2-x(1-x)m_\pi^2]/\Lambda ^2\Big) 
\label{fpi}
\ee
as well as the kaon decay constant 
\be
f_K^2= \frac 14 (m+m_s)^2 \frac {N_c}{4\pi ^2} \int _0^1 dx
\Gamma\Big(0,[xm^2+(1-x)m_s^2-x(1-x)m_K^2]/\Lambda ^2\Big) .
\label{fK}
\ee
Using eqs (\ref{BSeq},\ref{fpi}) the four parameters of the model
(\ref{NJL}), the coupling constant $G_{\rm NJL}$, the cutoff 
$\Lambda $ and the
two current masses $m^0$ and $m_s^0$ may be determined. The current
quark masses are obtained from the pion and kaon masses, 
$m_\pi=135$ MeV and $m_K=495$ MeV. Fixing now the pion decay constant
$f_\pi = 93$ MeV yields too small a value for the kaon decay constant, see 
table \ref{tab_1}. On the other hand, requiring $f_K=113$ MeV leaves us with
too large a value for $f_\pi$.\footnote{Determining $f_\pi$ or $f_K$
fixes the ratio $\Lambda /m$. This leaves one adjustable parameter,
\eg the coupling constant $G_{\rm NJL}$. However, as $G_{\rm NJL}$ is 
not very transparent we will use the gap equations (\ref{conmass}) to 
re--express it in terms of the up constituent mass $m$.}
In the following we will exclusively employ parameter sets which
reproduce the physical value of $f_\pi$.

The expressions for the decay constants are only logarithmically
divergent, which in a Pauli--Villars scheme could be cured by
a single subtraction. For the full three flavor model this, however,
would not be sufficient because the pertinent treatment of flavor
symmetry breaking also requires the solution to the quadratically
divergent gap--equation (\ref{conmass}). In ref \cite{Ka93} the sizable
effects of the second subtraction are discussed for meson observables 
which can be made finite by a single subtraction.

\begin{table}
\caption{{\sf \label{tab_1}
The up and strange quark constituent and current masses, the cutoff
and the pion and the kaon decay constants for the parameters used 
later. Results are taken from ref \protect\cite{We93}.}}
~
\newline
\centerline{
\begin{tabular}{c|cc|c|cc}
$m$ (MeV) & $m_s$ (MeV) & $m^0_s/m^0_u$ 
& $\Lambda$ (MeV) & $f_\pi$ (MeV) & $f_K$ (MeV) \\
\hline
350&577&23.5&641&93.0 &104.4\\
400&613&22.8&631&93.0 &100.3\\
450&650&22.4&633&93.0 & 97.4\\
\hline
350&575&24.3&698& 99.3&113.0\\
400&610&23.9&707&103.0&113.0\\
450&647&23.6&719&105.7&113.0\\
\end{tabular}}
\end{table}

For the investigation of the baryon sector we constrain the meson 
fields to the chiral circle, {\it i.e.} we replace $\Sigma$ by 
its VEV $\langle \Sigma \rangle$ in eq (\ref{defm}). 
Hence our NJL model soliton has the more complicated structure 
\begin{equation}
\label{collfield}
 M(\vec x) = \xi_0 (\vec x) \Big\langle \Sigma \Big\rangle \xi_0 (\vec x)
\end{equation}
with
\begin{equation}
\label{hedgehog}
\xi_0(\vec x) =  \left( \begin{array}{cc}
                  \exp{[i \Theta(r) \hat{r}\cdot\vec{\tau}/2]} & 0 \\
                  0 & 1 \\
                  \end{array} \right)
\end{equation}
containing the hedgehog {\it ansatz}. The computation of the 
classical energy associated with the unit baryon number sector
of this field configuration is standard \cite{Re89,Al96}
\be
E[\Theta]-E[\Theta=0]&=&
\frac{N_C}{2}\epsilon_{\rm V}
\left(1+{\rm sign}(\epsilon_{\rm V})\right)
+\frac{N_C}{2}\int^\infty_{1/\Lambda^2}
\frac{ds}{\sqrt{4\pi s^3}}\sum_\nu{\rm exp}
\left(-s\epsilon_\nu^2\right)
\nonumber \\* && \hspace{4cm} +
m_\pi^2 f_\pi^2\int d^3r  \left[1-{\rm cos}\Theta(r)\right] ,
\label{efunct}
\ee
where $\epsilon_\mu$ are the eigenvalues of the single particle
Dirac Hamiltonian 
\be
h_0=\vec{\alpha}\cdot\vec{p}+m\, \beta \,
{\rm exp}\Big(i\gamma_5\vec{\tau} \cdot \hat{r} 
\Theta(r)\Big) \hat{T}
+m_s\, \beta \hat{S}
\label{hamil}
\ee
In particular $\epsilon_{\rm V}$ refers to the valence quark level 
which is strongly bound in background of the hedgehog.  From the 
appearance of the strange and non--strange projectors 
$\hat{S}={\rm diag}_{\rm fl}(0,0,1)$ and
$\hat{T}={\rm diag}_{\rm fl}(1,1,0)$ we observe that the strange quarks
are not effected by the hedgehog field and hence obey a free Dirac equation.
The profile of the chiral angle $\Theta(r)$ is determined by self--consistently
minimizing the energy functional (\ref{efunct}) \cite{Re88,Al94a}.

In order to generate states with baryon quantum numbers we apply the 
collective approach. In this approach time dependent coordinates
$A(t)\in SU(3)$ which parameterize the orientation of the soliton in 
flavor space are introduced to approximate the unknown time dependent
solution to the full equations of motion, {\it i.e.}
\be
\xi(\vec{x},t)=A(t)\xi_0(\vec{x})A^\dagger(t)\, .
\label{colrot}
\ee
The quantum mechanical treatment of these {\it would--be} zero--modes
restores flavor covariance and due to the hedgehog {\it ansatz} 
also rotational covariance. Here we also recognize the advantage of
the parameterization (\ref{defm}), as it avoids the time--dependent 
rotation of the flavor variant vacuum configuration, $\xi\equiv1$.
A slightly different parameterization for the flavor  
rotating meson fields is discussed in ref \cite{Bl93}.

We now have to evaluate the action for the field configuration 
(\ref{colrot}). This is most conveniently achieved by transforming
to the flavor rotating frame $q^\prime=A(t)q$ \cite{Re89}. This 
transformation induces two additional pieces in the single particle 
Dirac Hamiltonian \cite{We92}
\be
h=h_0+h_{\rm rot}+h_{\rm SB}
\label{hmod}
\ee
which are due to the time dependence of the collective rotations
\be
h_{\rm rot}=-iA(t)^{\dagger} \dot{A}(t) = 
\sum_{a=1}^8 \Omega^a \frac{\lambda^a}{2}.
\label{hrot}
\ee
and the flavor symmetry breaking parameterized by the different
constituent quark masses
\be
h_{\rm SB}={\cal T}\beta\frac{m-m_s}{\sqrt{3}} 
\Big(D_{8i}\lambda_i+D_{8\alpha}\lambda_\alpha+(D_{88}-1)\lambda_8\Big)
{\cal T}^{\dagger}\, .
\label{hsb}
\ee 
For convenience the angular velocities $\Omega_a$ and the chiral 
rotation ${\cal T}=\cos{\frac{\Theta}{2}}-i\gamma_5\vec{\tau}\cdot\hat{r}
\sin{\frac{\Theta}{2}}$ have been introduced. Furthermore the 
various isospin invariant components in $h_{\rm SB}$ have been made
explicit\footnote{Throughout this paper we adopt the summation conventions 
$i,j,k,...=1,2,3$ and $\alpha,\beta,...=4,..,7$.}. The modifications
of the single particle Dirac Hamiltonian will be treated perturbatively 
since they are of sub--leading order either in $1/N_C$ or in $(m-m_s)/m$. 
Details of this calculation, which results in a Hamiltonian for the 
collective coordinates, may be traced from the literature \cite{We92}. 
The canonical momenta $R_a$ appearing in this Hamiltonian are related 
to the angular velocities
\be
R_a=\cases{-(\alpha^2 \Omega_a+\alpha_1 D_{8a}), & a=1,2,3\cr
           -(\beta^2\Omega_a+\beta_1 D_{8a}), & a=4,..,7\cr
            \frac{N_CB}{2\sqrt{3}}, & a=8}\quad .
\label{rgen}
\ee
The inertia parameters $\alpha^2,..,\beta_1$ are functionals of the
soliton profile, again for explicit expressions we refer to the literature
\cite{Al96}.
Furthermore $D_{ab}=(1/2){\rm tr}(\lambda_a A \lambda_b A^\dagger)$ 
denote the adjoint representation of the collective rotations. They 
appear in eq (\ref{rgen}) as a consequence of flavor symmetry breaking, 
{\it i.e.} $\alpha_1$ and $\beta_1$ are due to $h_{\rm SB}$ and hence
proportional to the mass difference $m-m_s$. In table (\ref{tab_2})
we list the numerical values for these inertia parameters.
\begin{table}[t]
\caption{{\sf \label{tab_2}
The fermion contribution to the inertia parameter in the 
quantization rule (\protect\ref{rgen}). The moments 
of inertia $\alpha^2$ and $\beta^2$ are given in ${\rm GeV}^{-1}$
while the coefficients $\alpha_1$ and $\beta_1$ are 
dimensionless. The up--quark constituent mass $m$ is in 
${\rm MeV}$. Note that when diagonalizing the collective
Hamiltonian a piece ($\sim 1/{\rm GeV}$) originating from 
induced kaon fields is augmented to the strange moment of 
inertia $\beta^2$ \protect\cite{We92}.}}
~
\newline
\centerline{\tenrm\smalllineskip
\begin{tabular}{c|ccc|ccc|ccc|ccc}
$m$&$\alpha^2_{\rm V}$ & $\alpha^2_{\rm vac}$ & $\alpha^2_{\rm tot}$ &
$\beta^2_{\rm V}$ & $\beta^2_{\rm vac}$ & $\beta^2_{\rm tot}$ &
$\alpha_{1\rm V}$ & $\alpha_{1\rm vac}$ & $\alpha_{1\rm tot}$ &
$\beta_{1\rm V}$ & $\beta_{1\rm vac}$ & $\beta_{1\rm tot}$\\
\hline
350 & 7.20 & 1.13 & 8.33 & 1.65 & 0.52 & 2.17 &
-0.925  & -0.004 & -0.929 & -0.203 & -0.003 & -0.206\\
400 & 4.55 & 1.26 & 5.81 & 1.33 & 0.52 & 1.85 &
-0.401  & -0.005 & -0.405 & -0.121 & -0.003 & -0.124\\
450 & 3.46 & 1.33 & 4.79 & 1.14 & 0.50 & 1.64 &
-0.232  & -0.006 & -0.238 & -0.081 & -0.004 & -0.085\\
\end{tabular}
}
\end{table}
In the process of quantization these canonical momenta are identified 
with the right generators of $SU(3)$. The resulting Hamiltonian $H(A,R)$ 
whose eigenstates are identified with the low--lying $\frac{1}{2}^+$ and 
$\frac{3}{2}^+$ baryons, also contains symmetry breaking terms.  
As a result of exactly diagonalizing $H(A,R)$ the $\frac{1}{2}^+$ 
baryons cease to be pure octet states, rather they
contain sizable admixture  of higher dimensional $SU(3)$
representations. Similarly the  $\frac{3}{2}^+$ baryons are no pure decuplet
states \cite{We96}.  
A major effect of these admxitures is to reduce the strange quark 
contributions to baryon properties, which are computed using the exact 
eigenfunctions of $H(A,R)$. This can easily be understood because due 
to the inclusion of symmetry breaking it is less probable for the heavier 
strange quarks to be excited. This effect will also become apparent in our 
results for the structure functions. Finally we should mention that the 
constraint for $R_8$ restricts the permissible $SU(3)$ representations 
to those containing only states with half--integer spins, 
{\it i.e.} fermions \cite{Wi83}.

\medskip

\leftline{\large\it 2b. The Hadronic Tensor in the NJL Chiral 
Soliton Model}

\medskip

The hadronic tensor which enters the cross--section 
of deep--inelastic scattering (DIS)
\be
W_{\mu \nu}^{ab}(P,q;S) = \frac{1}{4\pi} 
\int d^4x e^{iq\cdot \xi} 
\Big\langle P,S\Big| [ J_{\mu}^{a}(\xi),J_{\nu}^{b \dagger}(0)] 
\Big| P,S \Big\rangle 
\label{hten1}
\ee
represents the defining quantity for the structure functions. 
They are certain combinations (see below) of the form factors
appearing in the Lorentz--covariant decomposition (omitting
parity violating contributions)
\be
W_{\mu \nu}^{ab}(P,q;S) 
& = & \left(-g_{\mu \nu} + \frac{q_{\mu} q_{\nu}}{q^2}\right)
M_N W_{1} (x,Q^2) 
\nonumber \\ & & 
+\left(P_{\mu} - q_{\mu}\frac{P\cdot q}{q^2}\right) 
\left(P_{\nu} - q_{\nu}\frac{P\cdot q}{q^2}\right)
\frac{1}{M_N} W_{2} (x,Q^2) 
\nonumber \\ & &
+i\epsilon_{\mu \nu \lambda \sigma} \frac{q^{\lambda} M_N}{P\cdot q}
\left(\left[g_1(x,Q^2)+g_2(x,Q^2)\right]S^{\sigma}
-\frac{q\cdot S}{q\cdot P} P^{\sigma}
    g_2(x,Q^2) \right)\, .
\label{hten1a}
\ee
In eqs (\ref{hten1}) and (\ref{hten1a}) the nucleon state is 
characterized by its momentum $P$ and spin--orientation $S$. In the
nucleon rest frame they are $P^\mu=(M_N,{\vec 0})$ and 
$S^\mu=(0,{\vec s})$. The Lorentz invariant quantities are the 
momentum transfer squared, $Q^2=-q^2$ and the Bjorken variable 
$x=Q^2/ 2 P\cdot q$ where $q$ denotes the momentum of the exchanged 
boson. This boson couples to the hadronic current $J_\mu^a$ with $a$ 
being a suitable flavor component\footnote{We omit the flavor indices 
of the form factors.}.  Here it is important to note that this 
hadronic current is completely determined by the symmetry currents 
of QCD. Hence the hadronic current can unambiguously be computed in 
any model which has the same symmetries as QCD. This model current then 
completely saturates the hadronic tensor and at this point there is no 
need to identify the degrees of freedom of QCD within the model. 

In the NJL model the interaction among the quarks does not contain any 
derivatives as can be seen from the defining equation (\ref{NJL}). Hence
the currents are formally identical to those of a free Dirac  
theory, {\it i.e.} $J_\mu^a ={\bar q}\gamma_\mu(\gamma_5)t^a q$ with $t^a$ 
being a flavor generator. We will concentrate on the leading twist 
pieces of the structure functions. These are obtained from the Bjorken 
limit $-q^2,2P\cdot q \to \infty $ with $x$ fixed\footnote{That is,
the higher order contributions in $1/Q^2$ are dropped. The
logarithmic dependence on $Q^2$ originating from perturbative
gluon emission to the nucleon will be discussed in section 5.}. 
In this limit
the computation of the hadronic tensor facilitates considerably since we
may adopt the free correlation function for the highly off--shell intermediate
quark propagating between the two boson insertions:
\be
W_{\mu \nu}^{ab}(P,q;S)&=&\frac{1}{4\pi}
\int d^4\xi\, e^{iq\cdot \xi}
\int \frac{d^4k}{(2\pi)^4}\, e^{ik\cdot \xi}\, 
{\rm sign}(k_0)\, 2\pi\, \delta(k^2)
\nonumber \\ && \hspace{1cm} \times
\Big\langle P,S\Big|{\bar q}(\xi)\gamma_\mu t^a 
\kslash t^b \gamma_\nu q(0)
-{\bar q}(0)\gamma_\nu t^b \kslash t^a 
\gamma_\mu q(\xi)\Big| P,S \Big\rangle \, .
\label{hten2}
\ee
Apparently the hadronic tensor acquires contributions from intermediate
quarks propagating in forward ($\xi\to 0$) and backward ($0\to\xi$) 
directions. In order to apply this formalism to the soliton 
configuration, which breaks translational invariance, we still need 
to generate states of good linear momentum. This is essentially done 
in the same way as the flavor--spin projection discussed previously 
by introducing a collective coordinate ${\vec R}$ describing the 
position of the classical soliton in coordinate space. As the nucleons in 
the {\it bra} and {\it ket} states have the same momenta this boils 
down to an additional integral in coordinates space yielding \cite{Ja75}
\begin{eqnarray}
W_{\mu \nu}^{ab}(P,q;S) & = &  \frac{M_N}{(2 \pi)^4}
\int dt\, d^3\xi_1\, d^3\xi_2\, \int d^4k\, {\rm sign}(k_0)\, \delta (k^2) 
e^{i (q_0+k_0)t - i(\vec{q}+\vec{k})\cdot(\vec{\xi}_1-\vec{\xi}_2)} 
\nonumber \\* && \hspace*{3em} 
\times\Big\langle N \Big| \bar{q}(\vec{\xi}_1,t)
\gamma_\mu t^a\kslash t^b \gamma_\nu q(\vec{\xi}_2,0)-
\bar{q}(\vec{\xi}_2,0) \gamma_\nu t^b 
\kslash t^a \gamma_\mu q(\vec{\xi}_1,t) 
\Big| N \Big\rangle \, .
\label{hten3}
\end{eqnarray}
Here, $\Big| N \Big\rangle$ is defined via $\Big\langle\vec{R}\Big| P,S
\Big\rangle = \sqrt{2E}\exp(i\vec{P}\cdot\vec{R})\Big|N\Big\rangle$.
Although this now is a simplified version of the hadronic tensor it is 
not well suited for the ongoing calculation in the bosonized NJL model. 
The main reason is that the matrix element to be computed is that of a 
bilocal and bilinear quark operator which appears as an ordinary product
rather than a time--ordered product. Only the latter would be suitable to
undergo the bosonization procedure which in turn could provide a 
rigorous regularization of the structure functions consistent with 
the meson sector of the model as well as other soliton calculations.
One possible way to achieve that goal would be to start from the
forward Compton amplitude whose absorptive part is identical
to the hadronic tensor. This amplitude is calculated as the matrix
element of the time--ordered product of the electromagnetic currents
and can hence straightforwardly be implemented in the functional
bosonization thereby implementing a unique regularization. Although first
results \cite{Wexx} in this direction are promising\footnote{Since the 
incomplete gamma--functions appearing in the proper--time regularization 
have various unphysical poles in the complex plane it is more appropriate
to employ the Pauli--Villars regularization when identifying the cuts
of the Compton amplitude \cite{Da95}. In contrast to earlier two flavor 
calculations of the structure functions, which only included a single 
subtraction \cite{Di96,Wa98,Po99}, the computation of the Compton 
amplitude allows one to consistently incorporate the two subtractions 
demanded by the quadratically divergent gap--equation.} 
we will presently avoid the ambiguities connected with regularizing  
structure functions by employing the valence quark approximation. This 
approximation consists of substituting the valence quark wave--function 
\be
q(\vec \xi,t)=\psi_{\rm V} (\vec \xi,t) = e^{-i \epsilon_{\rm V} t} A(t) 
\left(\Psi_{\rm V} (\vec \xi)
+\sum_{\nu \neq \rm V} \Psi_{\nu} (\vec \xi)\,
\frac{\langle \nu | h_{\rm rot}+h_{\rm SB} |{\rm V} \rangle}
{\epsilon_{\rm V} - \epsilon_{\nu}} \right)
\label{valwfct}
\ee
in the above expression for the hadronic tensor (\ref{hten3}). The 
main arguments to consider this valence quark approximation to be 
reliable are the facts that the valence quark contribution does not 
undergo regularization in any event and the observation that this level 
dominates the results for the static properties. The axial properties 
are actually almost saturated by the valence quark contribution \cite{Al96}.
It is important to note that this valence level should not be confused 
with the valence quark in the parton model description of the structure 
functions. Here it rather refers to the distinct quark state which is 
strongly bound in the soliton background.

Note that both the rotational and symmetry breaking corrections have 
been included. This, of course, is necessary in order to disentangle 
various flavor contributions. Consistent $1/N_C$ counting makes it
necessary to consider the time-dependence of $A$\cite{Po99}. This additional
bilocality in the time coordinates is treated in an $1/N_C$ expansion 
employing the quantization rules (\ref{rgen})
\be
A(t)&=&A(t_0)+(t-t_0)\frac{dA(t)}{dt}\Big|_{t=t_0} 
+{\cal O}\left(\frac{1}{N_C^2}\right)
\nonumber \\
&=&A(t_0)+\frac{i}{2}(t-t_0)A(t_0)\sum_{a=1}^8\lambda^a\Omega_a(t_0)
+\ldots\, ,
\label{biloct}
\ee
where the ellipsis refer to higher order contributions in the angular
velocities which according to eq (\ref{rgen}) correspond to higher
orders in $1/N_C$ and/or flavor symmetry breaking. At this point 
we recognize that the computation of nucleon matrix elements 
associated with the expansion (\ref{biloct}) will also demand 
the matrix element of $\Omega_8$. This component of the angular
velocity is not provided by the quantization prescription (\ref{rgen}). 
We obtain the relevant information from 
\be
\Omega_8=-i{\rm tr}\left(\lambda^8 A(t)^\dagger {\dot A}(t)\right)
={\rm tr}\left(\lambda^8 A(t)[H,A(t)]\right)\, ,
\label{evom8}
\ee
with $H=H(A,R)$ being the Hamilton operator for the collective 
coordinates, its eigenstates are the baryons, {\it cf.} the
discussion after eq (\ref{rgen}). Employing the defining equation 
for right generators, {\it i.e.} $[A,R_a]=A\lambda_a/2$ yields
\be
\langle\Omega_8\rangle = 
\frac{\sqrt{3}}{4 \alpha^2} - \frac{1}{2\sqrt{3} \beta^2} \, .
\label{omega8}
\ee
It is straightforward to verify that the application of the same 
procedure to $\Omega_1,..,\Omega_7$ is consistent with the 
quantization rules (\ref{rgen}).

\bigskip
\leftline{\Large\it 3. Unpolarized Nucleon Structure Functions}
\medskip

We have now collected all ingredients necessary to apply the 
valence quark approximation to the computation of the hadronic tensor 
in the NJL chiral soliton model. In a first step we project the 
unpolarized structure function $f_1(x)$ from $W_{\mu\nu}$
by contracting it appropriately with
\be
\label{f1pro}
 \Lambda_{f_1}^{\mu \nu}= \frac{1}{2}(- g^{\mu \nu} +
 \frac{\eta}{M_N^2} P^{\mu} P^{\nu} ) 
\quad {\rm where}\quad
\eta = \left(1+ \frac{P\cdot q}{2xM_N^2} \right)^{-1}\, .
\ee
In the Bjorken limit, which in the nucleon rest frame
$P^\mu=(M_N,\vec{0}\,)^\mu$ is implemented via
\be
\label{bjl}
q_0 = |\vec{q}| - x M_N 
\quad {\rm with} \quad
|\vec{q}| \rightarrow \infty\, ,
\ee
the leading twist pieces of the structure functions are extracted. In 
this limit this projector becomes as simple as $-g^{\mu\nu}/2$  when
contracted with the hadronic tensor in the form of eq (\ref{hten3}). 
In this limit the evaluation of the contribution due to the bilocality 
in the time--coordinate (\ref{biloct}) also becomes feasible. The 
linear $t$--dependence is treated as a derivative with respect to 
$q^0$. The $q^0$ dependence of the structure functions occurs via 
the Bjorken variable $x$. Hence this derivative is replaced by 
$\frac{-1}{M_N}\frac{d}{dx}$. Accordingly it is convenient to separate 
the unpolarized structure function $f_1$ into two contributions:
\be
 f_1(x) = \tilde{f}_1(x) + \Delta f_1(x)
\ee
with $\Delta f_1(x)$ containing the piece due to bilocality 
in the collective coordinates (\ref{biloct}). By definition 
this contribution is already linear in the angular velocities, 
and hence of sub--leading order. Consequently the structure
function  
$\Delta f_1(x)$ is completely given in terms of the classical fields, 
{\it i.e.} by omitting the cranking correction in eq (\ref{valwfct}). 

It is henceforth useful to define the following Fourier--transformations
\be
\tilde{\psi}_{\rm V}(\vec{p})&=&
\int \frac{d^3 \xi}{4\pi}\left(\Psi_{\rm V} (\vec \xi)
+\sum_{\nu \neq \rm V} \Psi_{\nu} (\vec \xi)\,
\frac{\langle \nu | h_{\rm rot}+h_{\rm SB} |{\rm V} \rangle}
{\epsilon_{\rm V} - \epsilon_{\nu}} \right)
{\rm exp}\left(i\vec{\xi}\cdot\vec{p}\right) 
\hspace{1.7cm}
\label{ft1}
\\ {\rm and} \hspace{2.8cm} && \nonumber \\
\tilde{\Psi}_{\rm V}(\vec{p})&=&
\int \frac{d^3 \xi}{4\pi}\, \Psi_{\rm V} (\vec \xi)\,
{\rm exp}\left(i\vec{\xi}\cdot\vec{p}\right)\, .
\label{ft2}
\ee
According to eq (\ref{hten3}) we put $\vec{p}=\vec{k}+\vec{q}$
together with the on--shell condition 
$k=|\vec{k}|=q_0\mp\epsilon_{\rm V}$ pertinent for the forward and 
backward contributions to the hadronic tensor. As in the Bjorken limit 
$q_0$ tends to infinity and $q_0-k$ remains finite, we may
change the integration measure to
\be
k^2 d\Omega_{\vec{k}} = p dp d\phi
\ee
with $p=|\vec{p}|$ and $\phi$ being the azimuth angle between 
$\vec{q}$ and $\vec{p}$. The lower limit of the $p$--integration is 
assumed when $\vec{k}$ and $\vec{q}$ are anti--parallel, {\it i.e.} 
$p^{\rm min}_\pm=|M_Nx\mp\epsilon_{\rm V}|=:M_N|x_\mp|$. Finally we 
make use of the single quark wave--function containing only 
finite momentum modes of the order $p\approx\epsilon_{\rm V}$. Hence 
the integral receives non--vanishing contributions only when 
$\hat{k}=-\hat{q}$ \cite{Ja75,We96}. With these preliminaries the 
unpolarized structure function $f_1$ becomes
\be
\tilde{f}_1^{ab} & = & N_C \frac{M_N}{\pi} \Big\langle N \Big| 
 \int_{M_N |x_{-}|}^{\infty} p 
dp d\phi\, \tilde{\psi}_{\rm V}^{\dagger} (\vec{p}_{-}) 
  (1 - \vec{\alpha}\cdot\hat{q})
A^{\dagger} {\cal O}^{ba} A \tilde{\psi}_{\rm V} (\vec{p}_{-}) 
\nonumber \\ & & \hspace*{5em} -\int_{M_N
|x_{+}|}^{\infty} p dp d\phi\, 
 \tilde{\psi}_{\rm V}^{\dagger} (\vec{p}_{+}) 
  (1 + \vec{\alpha}\cdot\hat{q}) A^{\dagger} {\cal O}^{ab} 
  A \tilde{\psi}_{\rm V}
(\vec{p}_{+}) \Big| N \Big\rangle \nonumber \\
 & = & \tilde{f}_{1,-}^{ab} - \tilde{f}_{1,+}^{ab} 
\label{tf1}\\
\Delta f_1^{ab} & = & - \frac{N_C}{2 \pi} \frac{d} {d x} 
  \Big\langle N \Big| \int_{M_N 
  |x_{-}|}^{\infty} p 
dp d\phi\, \tilde{\Psi}_{\rm V}^{\dagger} (\vec{p}_{-}) 
  (1 - \vec{\alpha}\cdot\hat{q})
A^{\dagger} {\cal O}^{ba} A \Omega_g \lambda^g
 \tilde{\Psi}_{\rm V}(\vec{p}_{-})
 \nonumber \\ & & \hspace*{5em}+\int_{M_N |x_{+}|}^{\infty} 
  p dp d\phi\, \tilde{\Psi}_{\rm V}^{\dagger} 
  (\vec{p}_{+}) (1 + \vec{\alpha}\cdot\hat{q}) 
\Omega_g \lambda^g A^{\dagger} {\cal O}^{ab} A 
  \tilde{\Psi}_{\rm V} (\vec{p}_{+}) \Big| N \Big\rangle \nonumber 
\label{df1}\\
 & = & \Delta f_{1,-}^{ab} + \Delta f_{1,+}^{ab} 
\ee
where we have also used that the valence quark wave--function 
carries positive parity, {\it i.e.} 
$\psi_{\rm V}(-\vec{p})=\gamma_0\psi_{\rm V}(\vec{p})$. 
The momentum ${\vec p}_\pm$ denotes the integrated momentum
with the polar angle fixed, ${\rm cos}\Theta^\pm=M_N x_\pm/p$.
The flavor structure of the integrand is now comprised in the 
matrices ${\cal O}^{ab} = t^a t^b$ which are diagonal for all 
processes in which we are interested
\be
{\cal O}^{ab} & = & 
\delta_0 {\bf 1} + \delta_3 \lambda_3 + \delta_8 \lambda_8  
\nonumber \\
{\cal O}^{ba} & = &
\delta_0^\prime {\bf 1} + \delta_3^\prime \lambda_3 +
\delta_8^\prime  \lambda_8\, .
\label{flmat}
\ee
These matrices are given by the physical process under 
consideration as this specifies the flavor matrices $t^a$ and $t^b$ 
in the hadronic tensors. Since we will ignore Cabbibo mixing, the two 
sets $(\delta_0,\delta_3,\delta_8)$ and 
$(\delta^\prime_0,\delta^\prime_3,\delta^\prime_8)$ differ at most 
by the sign of $\delta_3$ and $\delta^\prime_3$.

We remark the change of sign in the backward moving quark contribution 
when going from $\tilde f_1$ to $\Delta f_1$. This stems from the
different expansion (\ref{biloct}) caused by exchanging the time
arguments of the collective coordinates.

According to the quantization rule (\ref{rgen}) the collective 
coordinates and their time derivatives $\Omega_a$ are elevated to 
non--commuting operators. Guided by symmetry requirements such as 
hermiticity, PCAC and G--parity we adopt the symmetric ordering 
whenever ambiguities occur\footnote{While G--parity invariance can 
be maintained for other orderings when going beyond the valence quark 
approximation this is not the case for PCAC \cite{Sch95}.}, {\it i.e.} 
$D_{ab}\Omega_c\rightarrow\{D_{ab},\Omega_c\}/2$.

For a consistent formulation of the valence quark approximation 
we consider the sum rules for isospin and strangeness of the 
nucleon. The former is known as the Adler sum rule and refers
to (anti)--neutrino scattering off the nucleon. For these processes
we require $t^a=(t^b)^\dagger=(\lambda^1\pm i\lambda^2)$. According
to eq (\ref{flmat}) the subtraction of neutrino and anti--neutrino 
scattering processes\footnote{Here we do not specify the 
boundaries of the integrals because they will be subject
to the boost into the infinite momentum frame, {\it cf.}
eq (\ref{RFtoIMF}).} 
\be
\int\frac{dx}{x}\left(f_2^{\bar{\nu}N}-f_2^{\nu N}\right)
\ee
corresponds to $\delta_3=-\delta^\prime_3=1$
with all other $\delta_a^{(\prime)}$s identical to zero. It is 
then straightforward to verify that

\be
\int dx \left(\tilde{f}_{1,-} - \tilde{f}_{1,+}\right) & = & 
\Big\langle N \Big| 2 I_3 \Big| N
\Big\rangle \label{Adler1} \\ 
\int dx \left(\Delta f_{1,-} + \Delta f_{1,+}\right) & = & 0 \, 
\ee
with $I_3 = D_{3i} R_i + D_{3 \alpha} R_{\alpha} + D_{38} R_{8}$.
However, for eq (\ref{Adler1}) to be valid, we have to restrict 
the inertial parameters appearing in the quantization rule (\ref{rgen})
to their dominating valence quark contribution. Of course, this 
is expected. The differences between using the total values
of these inertial parameters and these valence quark contributions
can be employed to estimate the uncertainties of the valence quark 
approximation.
Within the same restriction also the strangeness sum rule 
for $t^a=t^b=(1-\sqrt{3}\lambda^8)/3$ holds as one easily 
verifies that (for $N_C=3$)
\be
\int dx \left(\tilde{f}_{1,-} + \tilde{f}_{1,+}\right) & = & 
\frac{-1}{2}\Big\langle N \Big| \hat{S} \Big| N
\Big\rangle \label{Adler2} \\ 
\int dx \left(\Delta f_{1,-} - \Delta f_{1,+}\right) & = & 0 \, 
\ee
with $\hat{S} = (2 \sqrt{3} /N_C)(L_8 - R_8)$.
An interesting sum rule for the unpolarized structure functions
is the so--called Gottfried sum rule \cite{Go67} which addresses
electron--nucleon scattering, 
\be
S_{\rm G}=\int \frac{dx}{x}\Big(f_2^{\rm ep}-f_2^{\rm en}\Big)\, .
\label{sg1}
\ee
The structure function $f_2$ is related to the form factor $W_2$ in 
the defining equation (\ref{hten1a}) by $f_2=-q^2 W_2/2M_Nx$. In the
Bjorken limit the Callan Gross relation $f_2=2xf_1$ is 
identically fulfilled. Using the electromagnetic 
combination
\be
{\cal O}^{ab}={\cal O}^{ba}=(\lambda_3/2+\lambda_8/2\sqrt{3})^2
=:{\cal O}^{\rm e.m.}
\label{oem}
\ee
we therefore find
\be
S_{\rm G}=2\frac{N_C}{3\pi} \int dx \Big(s_G(x)+\triangle s_G(x)\Big)
\label{sg2}
\ee
with
\be
s_G(x)&=&M_N\Big\langle N \Big| \int_{M_N
  |x_{-}|}^{\infty} p
dp d\phi\, \tilde{\psi}_{\rm V}^{\dagger} (\vec{p}_{-})
  (1 - \vec{\alpha}\cdot\hat{q})
A^{\dagger} \lambda_3 A \tilde{\psi}_{\rm V} (\vec{p}_{-})
\nonumber \\ & & \hspace*{5em} -\int_{M_N
|x_{+}|}^{\infty} p dp d\phi\,
 \tilde{\psi}_{\rm V}^{\dagger} (\vec{p}_{+})
  (1 + \vec{\alpha}\cdot\hat{q}) A^{\dagger} \lambda_3
  A \tilde{\psi}_{\rm V}
(\vec{p}_{+}) \Big| N \Big\rangle \nonumber
\ee
and
\be
\triangle s_G(x)&=&-\frac{1}{2}\frac{d}{dx}
\Big\langle N \Big| \int_{M_N
  |x_{-}|}^{\infty} p
dp d\phi\, \tilde{\Psi}_{\rm V}^{\dagger} (\vec{p}_{-})
  (1 - \vec{\alpha}\cdot\hat{q})
A^{\dagger} \lambda_3 A \lambda^g \Omega_g
 \tilde{\Psi}_{\rm V}(\vec{p}_{-})
 \nonumber \\ & & \hspace*{5em}+\int_{M_N |x_{+}|}^{\infty} 
p dp d\phi\, \tilde{\Psi}_{\rm V}^{\dagger} (\vec{p}_{+})
  (1 + \vec{\alpha}\cdot\hat{q}) 
  \lambda^g \Omega_g A^{\dagger} \lambda_3 A 
  \tilde{\Psi}_{\rm V} (\vec{p}_{+}) \Big| N \Big\rangle\, . 
\nonumber
\ee
In the parton model language the assumption of a sea quark 
distribution which is identical for all flavors yields the historic 
value $S_G=1/3$. Recent experiments by the NMC group indicate a 
sizable deviation $S_G=0.235\pm0.026$ \cite{NMC94}. This has
often been interpreted as an isospin violation. However, this
is not the case because a non--vanishing lower component of the
quark wave--function $\psi_{\rm V}$ is sufficient to provide the
desired deviation from the historic value even for identical
up and down quark masses \cite{We96,Wa98}. Since the quarks 
experience a mesonic background field this effect may
be attributed to pion clouds \cite{Wa92}.

\bigskip
\leftline{\Large\it 4. Polarized Structure Functions}
\medskip 

The projector suitable to extract the polarized twist--2 structure 
function $g_1$ from the hadronic tensor reads \cite{An95}
\be
\label{g1projektor}
\Lambda_{g_1}^{\mu \nu}&=& \frac{2}{b} \left(2 P\cdot q x S_\rho +
\frac{1}{q\cdot S} \left\{\left(q\cdot S\right)^2 - 
\left(\frac{P\cdot q}{M_N}\right)^2 \right\} q_\rho\right) P_\sigma
\epsilon^{\mu \nu \rho\sigma} 
\label{g1pro} \\
{\rm with} \qquad
b&=&-4M_N\left\{\left(\frac{P\cdot q}{M_N}\right)^2
+2P\cdot q x -\left(q\cdot S\right)^2\right\}
\quad {\rm and} \quad \vec{s} \  \| \ \vec{q}\, .
\ee
In the Bjorken limit (\ref{bjl}) its contraction with 
the hadronic tensor (\ref{hten3}) is equivalent to 
the multiplication of
\be
\Lambda_{g_1,{\rm Bj}}^{\mu\nu}=\frac{1}{2M_N}
\epsilon^{\mu\nu\lambda\tau}\frac{q_\lambda P_\tau}{q\cdot S} \, .
\label{g1prbj}
\ee
Taking into account that for $g_1$ the nucleon spin $\vec s$ should 
be oriented along the photon momentum $\vec q$ which is accomplished
by the choice 
\be
{\hat q} = {\hat s} = {\hat e}_z
\ee
we obtain for this polarized structure function
by repeating the computation of the previous section 
\be
\tilde{g}_1^{ab} & = & N_C \frac{M_N}{\pi} 
  \Big\langle N ,{\hat s = {\hat e}_z}\Big| \int_{M_N
  |x_{+}|}^{\infty} p
dp d\phi\, \tilde{\psi}_{\rm V}^{\dagger} (\vec{p}_{+})
  (1+\alpha_3)\gamma_5
A^{\dagger} {\cal O}^{ab} A \tilde{\psi}_{\rm V} (\vec{p}_{+})
\nonumber \\* & & \hspace*{5em} -\int_{M_N
|x_{-}|}^{\infty} p dp d\phi\,
 \tilde{\psi}_{\rm V}^{\dagger} (\vec{p}_{-})
  (1-\alpha_3)\gamma_5 A^{\dagger} {\cal O}^{ba}
  A \tilde{\psi}_{\rm V}
(\vec{p}_{-}) \Big| N,{\hat s = {\hat e}_z}\Big\rangle \nonumber \\
 & = & \tilde{g}_{1,+}^{ab} - \tilde{g}_{1,-}^{ab}
\label{tg1}\\
\Delta g_1^{ab} & = &  \frac{N_C}{2 \pi} \frac{d} {d x} 
  \Big\langle N,{\hat s = {\hat e}_z}\Big| \int_{M_N
  |x_{+}|}^{\infty} p
dp d\phi\, \tilde{\Psi}_{\rm V}^{\dagger} (\vec{p}_{+})
  (1+\alpha_3)\gamma_5
\lambda^g \Omega_g A^{\dagger} {\cal O}^{ab} A
 \tilde{\Psi}_{\rm V}(\vec{p}_{+})
 \nonumber \\ & & \hspace*{5em}+\int_{M_N
|x_{-}|}^{\infty} p dp d\phi\, \tilde{\Psi}_{\rm V}^{\dagger} 
  (\vec{p}_{-}) (1 -\alpha_3)\gamma_5 A^{\dagger} 
 {\cal O}^{ba} A \Omega_g \lambda^g
  \tilde{\Psi}_{\rm V} (\vec{p}_{-}) \Big| N, {\hat s = {\hat e}_z}
 \Big\rangle \nonumber
\label{dg1}\\
 & = & \Delta g_{1,+}^{ab} + \Delta g_{1,-}^{ab}\, .
\ee
Again we have separated the contributions arising from 
time dependence of the collective coordinates (\ref{biloct})
\be
g_1^{ab}=\tilde{g}_1^{ab} +\Delta g_1^{ab}\, .
\label{g1tot}
\ee
Of course, the flavor indices have to be chosen according to
the considered process. We would like to remark that similar
to the case of the unpolarized structure functions the 
decomposition (\ref{g1tot}) does not directly correspond
to an expansion in $1/N_C$ and flavor symmetry breaking
because the valence quark wave--function $\psi_{\rm V}$
contains sub--leading orders as well, {\it cf.} eq
(\ref{valwfct}).

For the case of the electro--magnetic interaction 
${\cal O}={\cal O}^{\rm e.m.}$, {\it cf.} eq (\ref{oem}),
the sum rules discussed below are straightforwardly verified 
in the valence quark approximation when the previous restrictions 
to the inertia parameters as well as the static properties
are applied. The Bjorken sum rule \cite{Bj66} is obtained
from the difference between the first moments of the proton and 
neutron polarized structure functions
\be
\Gamma_1^{(p)}-\Gamma_1^{(n)}=
\int dx \left(g_1^{(p)}-g_1^{(n)}\right)
=\frac{1}{6} g_A
\label{bjsr}
\ee
with $g_A$ being the axial (isovector) charge of the nucleon measured 
in neutron beta--decay. In flavor SU(3) the isosinglet charge contains
octet as well as singlet pieces. The sum rule for the sum of the 
first moments is obtained as 
\be
\Gamma_1^{(p)}+\Gamma_1^{(n)}=
\int dx \left(g_1^{(p)}+g_1^{(n)}\right)
=\frac{5}{18}\left(\triangle u +\triangle d\right)
+\frac{1}{9}\triangle s
\label{axialsinglet}
\ee
where $\triangle q = \langle N, {\hat s = {\hat e}_z}
|{\bar q}\gamma_3\gamma_5 q| N, {\hat s = {\hat e}_z}\rangle$\, 
refers to the nucleon matrix element of the axial current
of flavor $q$. In this notation we have 
$g_A = \triangle u -\triangle d$.

The flavor singlet combination 
\be
\triangle u +\triangle d +\triangle s =\Sigma
\label{sigma}
\ee
corresponds to twice that part of the nucleon spin which 
is carried by the quarks, {\it i.e.} $\Sigma=2S_q$.
Its study has lead to the proton spin puzzle as the EMC \cite{EMC},
SMC \cite{SMC} and SLAC \cite{SLAC,SLAC98}
measurements combined with the assumption of flavor symmetry and data 
from semileptonic hyperon decays indicated that $\Sigma$ was 
unexpectedly small \cite{Br88}. In addition that analysis yielded a 
large polarization of the strange sea in the nucleon, {\it i.e.} a
large $\triangle s/\triangle d$. However, it was also observed
that this ratio is  quite sensitive to the assumption of SU(3) flavor
symmetry, while the smallness of $\Sigma$ remains almost unchanged
once this assumption is waived \cite{Pa89}. Our results for 
$\triangle q$ have already been presented in \cite{Sch98}. In this
context it should be noted that (as long as 
$\delta_3 = \delta_3^\prime$) the bilocal corrections do 
not alter $\triangle q$, {\it i.e.}
$\int dx (\Delta g_{1,+}^{ab} + \Delta g_{1,-}^{ab}) = 0$.

\bigskip 
\leftline{\Large\it 5. Projection and DGLAP--Evolution}
\medskip 
The structure functions of the previous sections are calculated 
in the nucleon ({\it i.e.} soliton) rest frame (RF). In the 
introduction we have already argued that we should transform these 
structure functions to the infinite momentum frame (IMF) in order 
to mitigate the effects originating from omitting the dynamical 
response\footnote{Hence the structure functions are (reference) 
frame--dependent quantities.} of the soliton on the transferred 
momentum. This boost causes a Lorentz contraction \cite{Ja80,Ga97}
\begin{equation}
\label{RFtoIMF}
f_{\rm IMF}(x) = \frac{\Theta(1-x)}{1-x} 
f_{\rm RF}\Big(-\ln{(1-x)}\Big)
\end{equation}
which constrains the structure functions to the interval
$x\in [0,1]$ hence providing proper support. The feature
that the structure functions vanish for $x\le1$ is not only
demanded by Lorentz covariance but also mandatory in order 
to include the logarithmic corrections in $Q^2$ employing the 
DGLAP evolution program \cite{Gr72,Al77} of perturbative QCD which 
otherwise contained ill--defined singularities. These corrections 
are incorporated by integrating differential equations ({\it cf.}
eqs (\ref{g1ns})--(\ref{ggs})) for the nucleon structure functions 
with respect to $t={\rm ln}(Q^2/\Lambda_{\rm QCD}^2)$. The 
assumption underlying the application of the QCD evolution program 
to the structure functions obtained in the present model (actually 
any model) is that the model approximates QCD at a low scale, 
$Q_0^2$. Hence at this scale the structure functions computed in 
the model and those of QCD should approximately be equal. This
low scale $Q_0^2$ represents the initial boundary 
$t_0={\rm ln}(Q_0^2/\Lambda_{\rm QCD}^2)$ for integrating the 
differential equations. It should be stressed that the lower 
scale $Q_0^2$ represents a new parameter being intrinsic to the model. 
We will fix $Q_0^2$ by demanding a best fit to the experimental data 
for the unpolarized structure function at the upper boundary of the 
$t$--integration. To be specific we will consider the linear
combination entering the Gottfried sum rule (\ref{sg2}).
As the DGLAP procedure relies on perturbative QCD we should
not take too small a value for $Q_0^2$ in order to stay 
in its range of validity. Of course, the upper boundary of 
the $t$--integration is set by the experiment to which the 
model results are compared.

Before actually integrating the DGLAP equations the flavor 
singlet (s) and non--singlet (ns) pieces of the structure functions 
have to be disentangled as the former may also contain gluon 
contributions. In the notation of eq (\ref{flmat}) the singlet 
component is denoted by $\delta_0$ while both $\delta_3$ and
$\delta_8$ are non--singlet. Formally the differential equations 
to be integrated read
\be
\frac{d f^{\rm (ns)}(x,t)}{dt}&=&
\frac{\alpha_{\rm QCD}(t)}{2\pi}\int_x^1 \frac{dy}{y}
P_{qq}\left(\frac{x}{y}\right) f^{\rm (ns)}(y,t)\ ,
\label{g1ns} \\ \nonumber \\
\frac{d f^{\rm (0)}(x,t)}{dt}&=&
\frac{\alpha_{\rm QCD}(t)}{2\pi}\int_x^1 \frac{dy}{y} \left\{
P_{qq}\left(\frac{x}{y}\right) f^{\rm (0)}(y,t)
+6P_{qg}\left(\frac{x}{y}\right) g(y,t) \right\}\ ,
\label{g1s} \\
\frac{d g(x,t)}{dt}&=&
\frac{\alpha_{\rm QCD}(t)}{2\pi}\int_x^1 \frac{dy}{y} \left\{
P_{gg}\left(\frac{x}{y}\right) g(y,t)
+P_{gq}\left(\frac{x}{y}\right) f^{\rm (0)}(y,t) \right\}\ .
\label{ggs}
\ee
Here the quantity $f$ symbolically represents both the unpolarized and 
polarized structure functions $f_1$ and $g_1$, respectively. Note that 
the splitting functions $P_{ij}$ which are listed in ref \cite{Al94} 
differ in these two cases. At this point we have to make a further 
assumption as the model does not give any information about the 
gluon content $g(x,t_0)$ in the nucleon which arises from quarks 
radiating and absorbing gluons and become {\it visible} only as the 
nucleon is probed with higher photon momenta\footnote{However,
gluonic correlations are already contained in the model which
can be concluded from the non--vanishing twist--3 structure function
${\bar g}_2$ \cite{We96}. The QCD equations of motion can be used to
show that the moments of ${\bar g}_2$ are matrix elements of the 
gluon field strength \cite{Sh82}. For a review on spin structure 
functions see ref \cite{Ja95}.}. We hence assume that at $Q_0^2$ the 
gluon distribution function for both the unpolarized and polarized 
structure functions vanish. This is not unmotivated as low--scale 
parameterizations of the nucleon distribution functions indicate that 
the gluon component is indeed small \cite{GRV,Gl98}. Hence the initial 
values of the integration are the model predictions of the nucleon 
structure functions together with $g(x,t_0)=0$. Note that this assumption
is irrelevant when determining $t_0$ from the unpolarized structure 
functions which enter the Gottfried sum rule (\ref{sg2}) because the
gluon distribution does not contribute to this isovector combination.
Having carried out the DGLAP integration to the scale $Q^2$, the 
singlet and non--singlet pieces are recombined to the physical flavor 
combinations. As a consequence of the evolution (\ref{ggs}) we find 
non--vanishing gluon contributions to the singlet structure function 
at any $Q^2>Q_0^2$.

In the numerical treatment we constrain the evolution equations 
to their leading order components in the QCD coupling 
$\alpha_{\rm QCD}(t)=4\pi/[(11-2N_f/3)\ln (Q^2/\Lambda^2_{\rm QCD})]$
not only for simplicity\footnote{From bag model calculations 
it is known that merely the variation of the model scale $Q_0^2$ is 
the major effect of taking the next to leading order correction into 
account \cite{St95}.} but in particular because a next to leading order 
calculation requires the identification of valence and sea 
quark distributions at the scale $Q_0^2$. This cannot be done in 
the present model as we merely identify the symmetries of QCD rather 
than the quark degrees of freedom. Note again that the valence quark 
level (\ref{valwfct}) is that of the NJL chiral soliton model. It is 
hence a constituent type quark degree of freedom and should not be 
confused with the valence distribution of the parton model. 
Actually the gap--equation (\ref{conmass}) causes the 
constituent quarks to contain not only current quarks
but also virtual pairs of current quarks and anti--quarks.

\bigskip
\leftline{\Large\it 6. Discussion of Numerical Results}
\medskip
We are now prepared to compare our model predictions to
the experimental data obtained for DIS at CERN and SLAC.

Within this model a reasonable description of the static 
properties of baryons is achieved for constituent quark
masses in the range $m=400\ldots450{\rm MeV}$ \cite{Al96}. 
We will henceforth consider the two limiting values for the 
only free parameter in the baryon sector.

We set the stage by discussing our results for the Gottfried sum rule 
(\ref{sg1}) in table \ref{tab_sg}. The value of this integral 
remains unchanged in the process of transforming to the IMF as 
well as the subsequent leading order DGLAP evolution. In presenting 
the results we have disentangled the two pieces associated 
to the integrand in eq (\ref{sg2}). Apparently the bilocal
corrections provide a sizable negative contribution; a result
which was already observed in the two flavor model \cite{Po99}. 
In table \ref{tab_sg} the numerical results for two different 
calculations are shown. Those labeled by `val' originate from
substituting only the valence quark piece of the inertia 
parameters when eliminating the angular velocities in favor
of SU(3) operators (\ref{rgen}) and matrix elements (\ref{omega8}).
In this case consistency of the valence quark approximation with the 
Adler sum rule is maintained \cite{We96}, see also section 3. The 
entry `val+vac' refers to the inclusion of the usually significantly 
smaller vacuum pieces in the inertia parameters. For nucleon 
properties which are obtained from matrix elements of vector currents 
the valence quark commonly contributes to about 80\% to the static 
properties \cite{Al96} which in the case of vector charges are 
evidently related to the inertia parameters. For axial properties 
the valence level more or less saturates the static properties 
\cite{Al96,Bl96}.
For those reasons we consider the valence quark approximation
to be most reliable when restricting to the `val' prescription in the 
case of the unpolarized structure functions but using the total
inertia parameters when later discussing the polarized structure
functions. After these general remarks we return to the Gottfried 
sum rule and observe that for $m=400{\rm MeV}$ a fair agreement
is obtained with the empirical value $0.235\pm0.026$ \cite{NMC94}.
Our result for $m=450{\rm MeV}$ is apparently on the low side. 
This may be due to the valence quark approximation getting worse
for larger coupling because a stronger binding of the valence quark
is joined by a more pronounced polarization of the vacuum.
\begin{table}[t]
\caption{{\sf \label{tab_sg}In this table we present our 
results for the Gottfried sum rule (\protect\ref{sg2}). 
The entries ``val'' and ``val+vac'' indicate whether 
only the valence quark pieces or the total values 
for the inertia parameters $\alpha^2$,..,$\beta_1$ 
are substituted in the quantization rules (\protect\ref{rgen}).
See also table \protect\ref{tab_1}.}}
\vspace{0.2cm}
\centerline{
\begin{tabular}{c|cc|cc|cc}
&\multicolumn{2}{c|}{$S_G$}&\multicolumn{2}{c|}{$\triangle S_G$} 
&\multicolumn{2}{c}{$S_G+\triangle S_G$}\\
\hline
$m$ {\footnotesize (MeV)} & val & val+vac 
& val & val+vac  & val & val+vac \\
\hline
400 & 0.29 & 0.22 & -0.07 & -0.05 & 0.22 & 0.17 \\
450 & 0.27 & 0.20 & -0.11 & -0.08 & 0.16 & 0.12 
\end{tabular} }
\end{table}

\begin{figure}[th]
\caption{{\sf \label{fig_f1gott}Numerical
results for the unpolarized structure functions entering the
Gottfried sum rule for $m=400{\rm MeV}$ (left panel) and
$m=450{\rm MeV}$ (right panel). Here the label RF indicates
the structure functions as calculated from the expressions
in the appendix, the transformation to the infinite momentum
frame (\protect\ref{RFtoIMF}) is indicated by IMF and finally
the structure functions resulting from the leading order (LO)
DGLAP evolution are represented by the solid line. The empirical
data are those for $(f_2^{\rm ep}-f_2^{\rm en})/2x$ \protect\cite{NMC94}.
Only the valence part of the moments of inertia enters the
quantization rule.}}
\vspace{0.5cm}
\centerline{\hspace{0cm}
\epsfig{figure=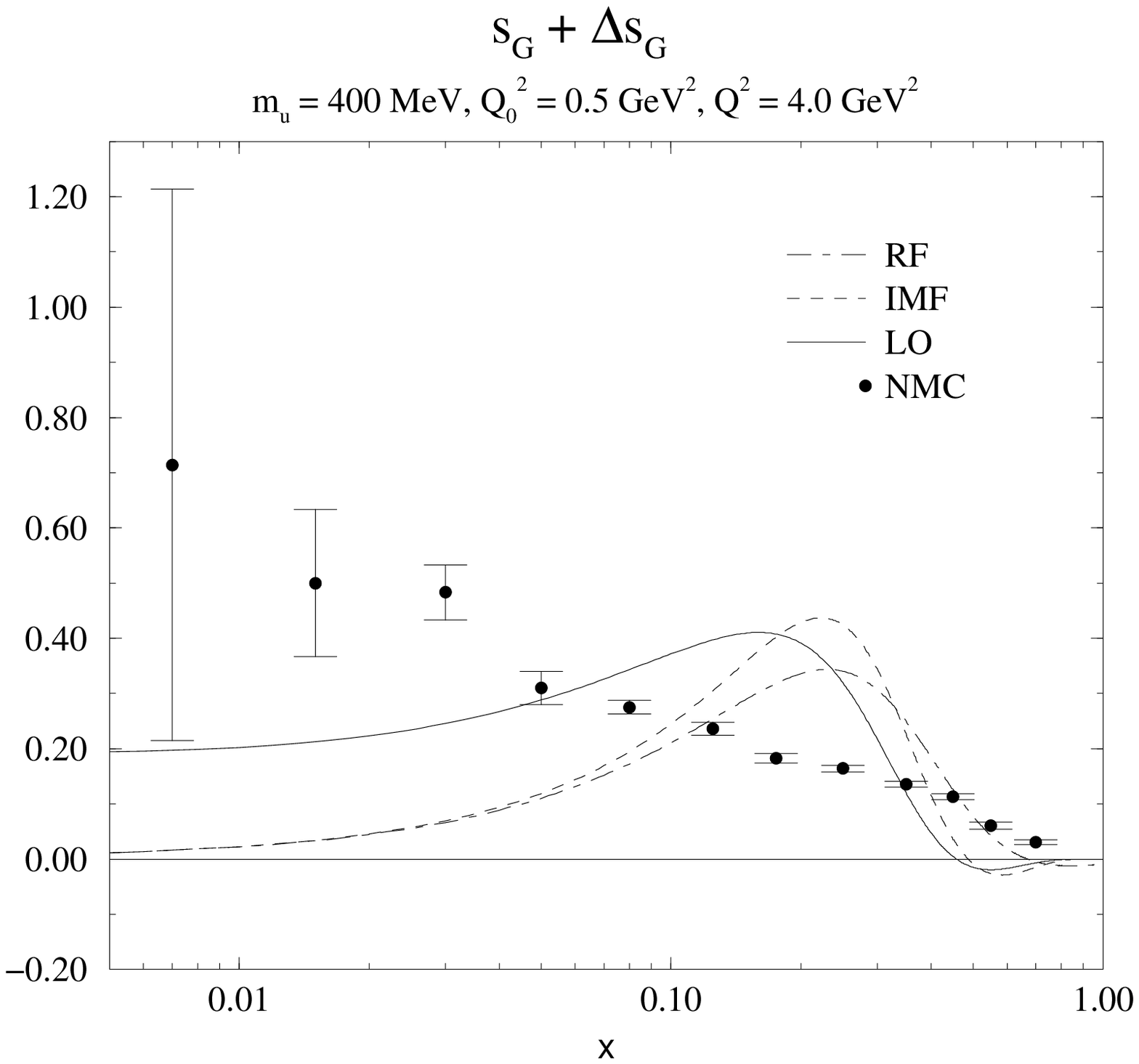,height=6cm,width=7.5cm}
\hspace{1cm}
\epsfig{figure=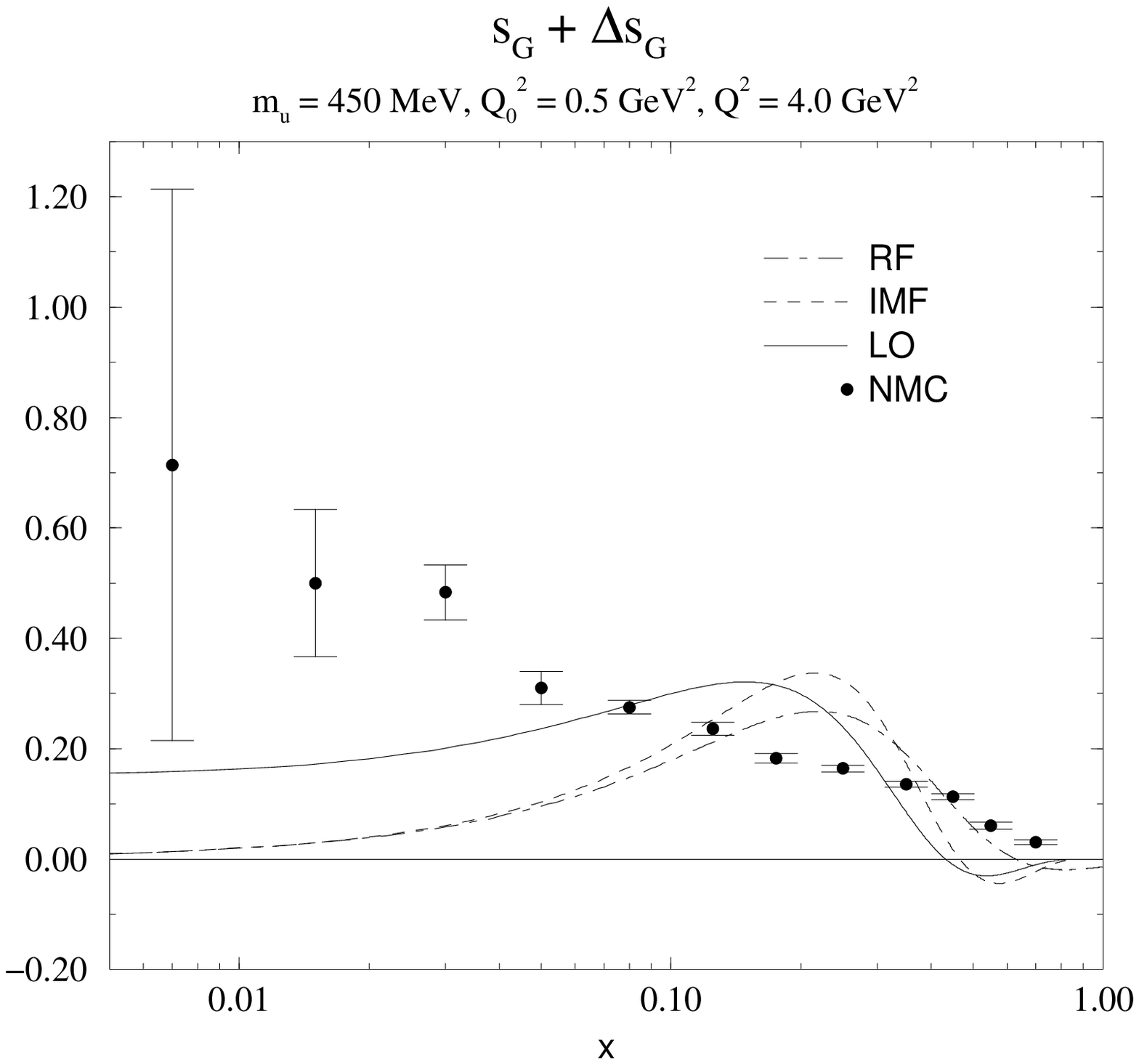,height=6cm,width=7.5cm}}
\end{figure}
We proceed by comparing the integrand of the Gottfried sum
rule, $s_G+\triangle s_G$ in eq (\ref{sg2}), with the 
empirical data, which we assume to obey the Callan Gross 
relation, $f_2=2xf_1$. For this comparison we apply the 
DGLAP evolution from the model scale $Q_0^2$ to the scale
$Q^2=4GeV^2$ at which the data are available. At $Q^2_0$ the
integrand of the DGLAP equations is taken to be the combination 
$s_G+\triangle s_G$ transformed to the IMF\footnote{Although the 
Callan Gross relation holds in our model for the structure functions
at the model scale $Q_0^2$ it does not necessarily hold after 
projection and evolution. Hence we consider it appropriate to
compare our results with the data obtained by applying the 
Callan Gross relation to the available data for $f_2$.}. 
By assimilating to the overall shape of the data we determine 
the model scale to be $Q_0^2\approx 0.5{\rm GeV}^2$, as shown
in figure \ref{fig_f1gott}. A smaller value for $Q_0^2$ might
even improve on the agreement as it lowers the structure
functions at large and moderate $x$ by simultaneously increasing 
it at small $x$. However, for too small a $Q_0^2$ one would have 
to go beyond leading order in the splitting functions. As already
discussed above this would require additional assumptions when 
identifying the model structure functions with those of QCD. From 
figure \ref{fig_f1gott} we also note that at $x\approx0.7$ the 
combination entering the Gottfried sum rule becomes slightly 
negative. This negative contribution is completely due to the
bilocal corrections $\triangle s_G$.

\begin{figure}[ht]
\caption{{\sf \label{fig_f1s}Model predictions for the unpolarized structure 
functions of the strange projector for two different values of the 
constituent up quark mass. Only the valence part of the moments of 
inertia enters the quantization rule.}}
\vspace{0.5cm}
\centerline{\hspace{0cm}
\epsfig{figure=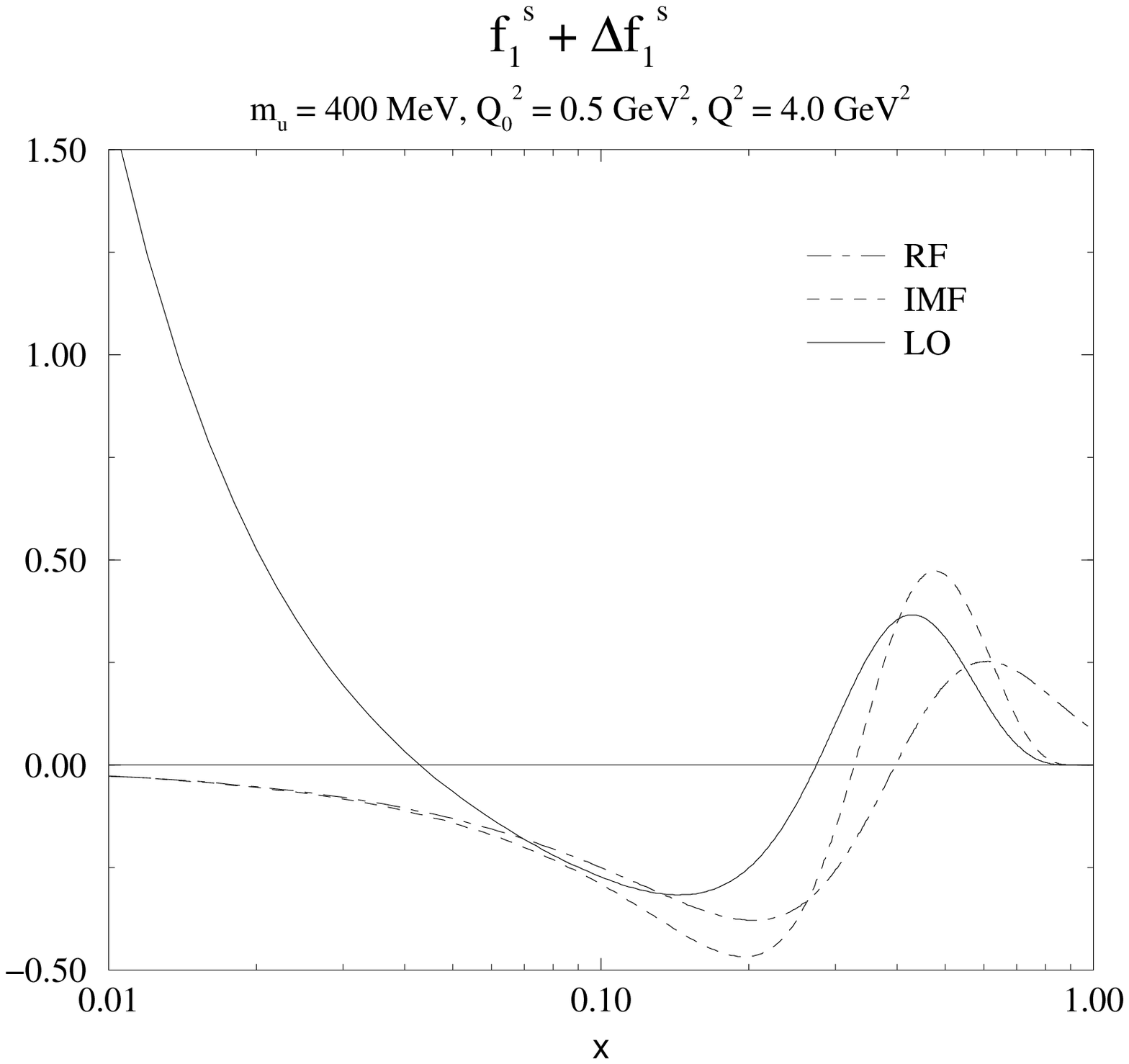,height=6cm,width=7.5cm}
\hspace{1cm}
\epsfig{figure=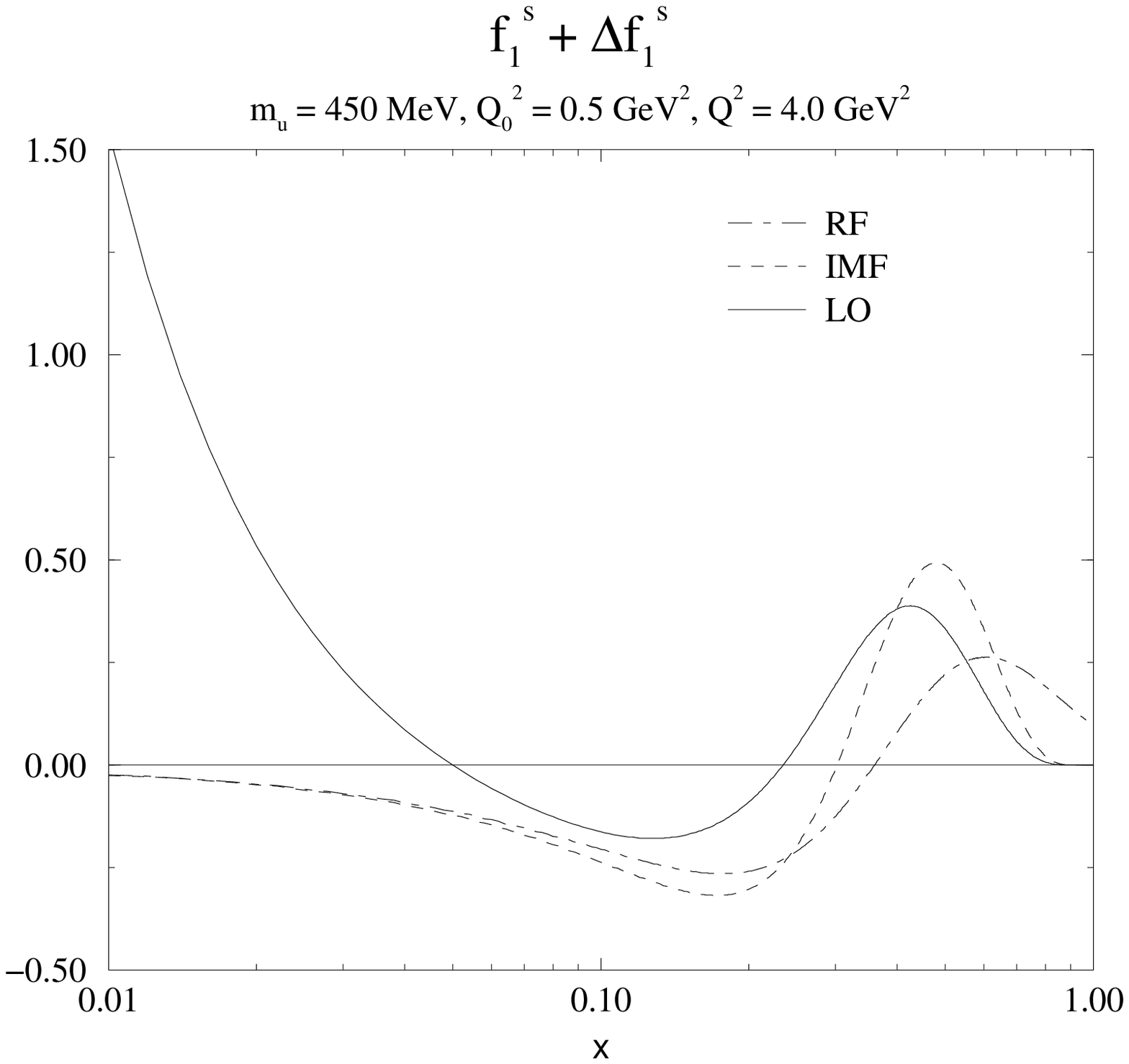,height=6cm,width=7.5cm}}
\end{figure}

Now we turn to an interesting quantity only accessible in 
the three flavor extension of the model: the contribution 
of strange degrees of freedom to the unpolarized structure
function, $f_1^s(x)$. Of course, a major motivation for the 
present study is the investigation of quantities like $f_1^s(x)$. 
This structure function is obtained by setting ${\cal O}^{ ab} 
= {\cal O}^{ba} = (1-\sqrt{3} \lambda_8)/3 $. Note that
the integral over $f_1^s(x)$ is not given by the strangeness 
charge discussed in eq (\ref{Adler2}) as for these two 
quantities the pieces stemming from the backward moving quarks 
have opposite signs. Our results for $f_1^s(x)$ are shown in 
figure \ref{fig_f1s}.

We observe that the structure function $f_1^s$ exhibits a pronounced 
maximum at $x\approx0.5$ after boosting the rest frame structure 
function to the infinite momentum frame. Na\"\i vely one expects 
the momentum distribution at the low energy scale to be peaked 
around $x=1/3$. However, the constituent mass of strange quark is 
about 50\% larger than that of the up or down quark, {\it cf.}
table \ref{tab_1}, which explains the shift of the maximum to a larger 
value in $x$. We also note that this maximum increases with the 
constituent mass $m$. This feature can be understood by recalling 
that this parameter actually measures the coupling of the non--strange
quarks to the soliton with the soliton then exciting the strange 
quarks. The DGLAP evolution only leads to a moderate smearing 
of the maximum. 

At lower $x$ the strangeness projected unpolarized 
structure function becomes negative which makes a probability 
interpretation in the spirit of the parton model impossible. This 
short--coming is most likely linked to the valence quark approximation
which not only omits contributions from the polarized vacuum (which 
still miss the formulation of a consistent regularization) 
as well as certain disconnected diagrams \cite{Ja75} which in the 
background of the soliton are not necessarily disconnected. 
Apparently the property of a negative piece in $f_1^s(x)$ gets 
mitigated at higher scales as the DGLAP evolution increases this
structure function at small $x$, it even might diverge as 
$x\to0$. We would like to remark that this divergence is
already present when the bilocal corrections $\triangle f_1$
are omitted \cite{Sch98a}.

\begin{figure}[bht]
\caption{{\sf \label{fig_g1ep}Numerical results for the polarized 
structure functions for electron--proton scattering for two 
different values of the constituent up--quark mass. 
The experimental values are taken from the SLAC E143 experiment 
\protect\cite{SLAC98}. The sum of the valence and the vacuum parts 
of the inertia parameters enter the quantization rule.}}
\vspace{0.5cm}
\centerline{\hspace{0cm}
\epsfig{figure=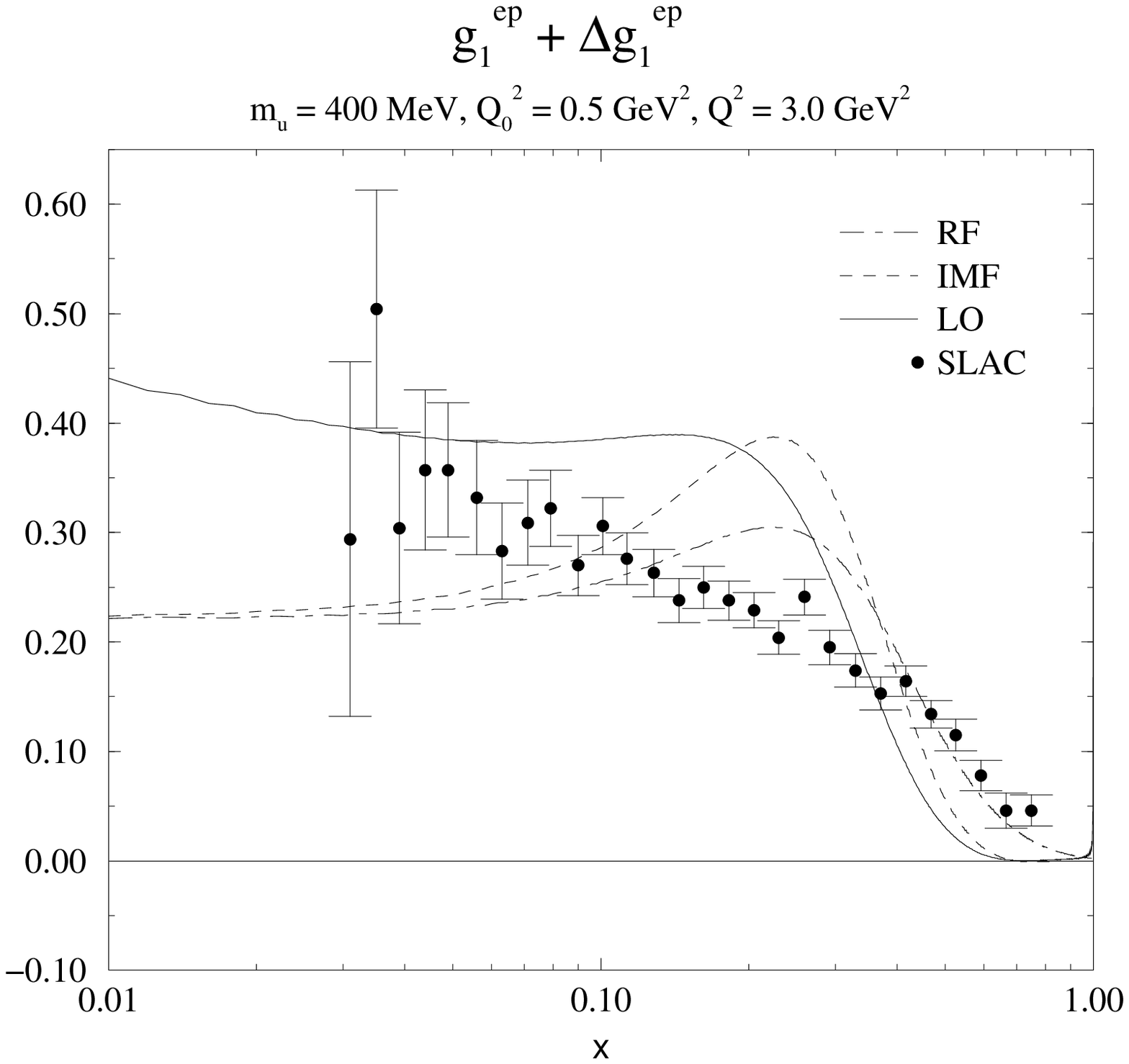,height=6cm,width=7.5cm}
\hspace{1cm}
\epsfig{figure=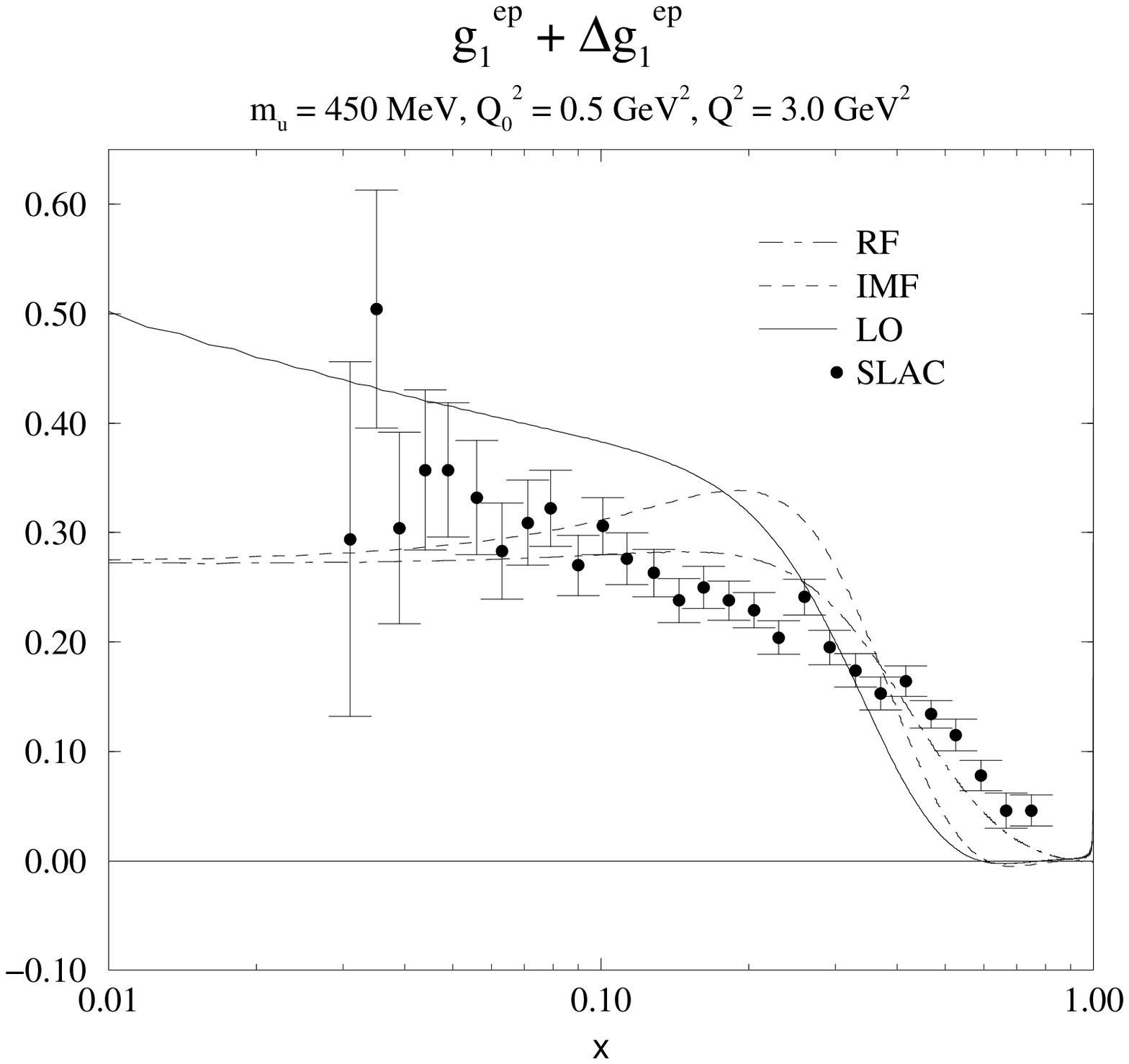,height=6cm,width=7.5cm}}
\end{figure}

\begin{figure}[th]
\caption{{\sf \label{fig_g1en}Same as figure \protect\ref{fig_g1ep}
for electron--neutron scattering.}}
\vspace{0.5cm}
\centerline{\hspace{0cm}
\epsfig{figure=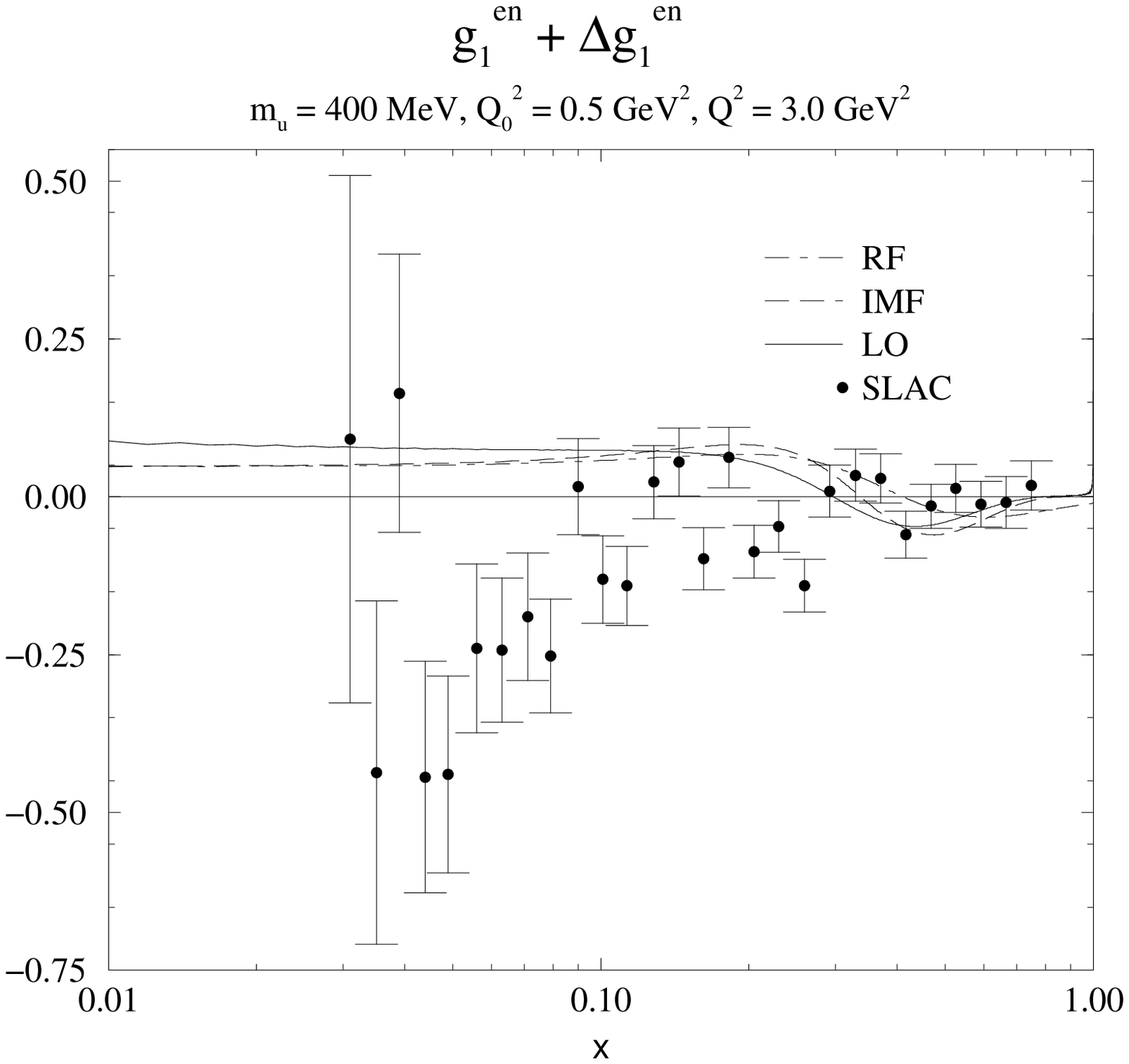,height=6cm,width=7.5cm}
\hspace{1cm}
\epsfig{figure=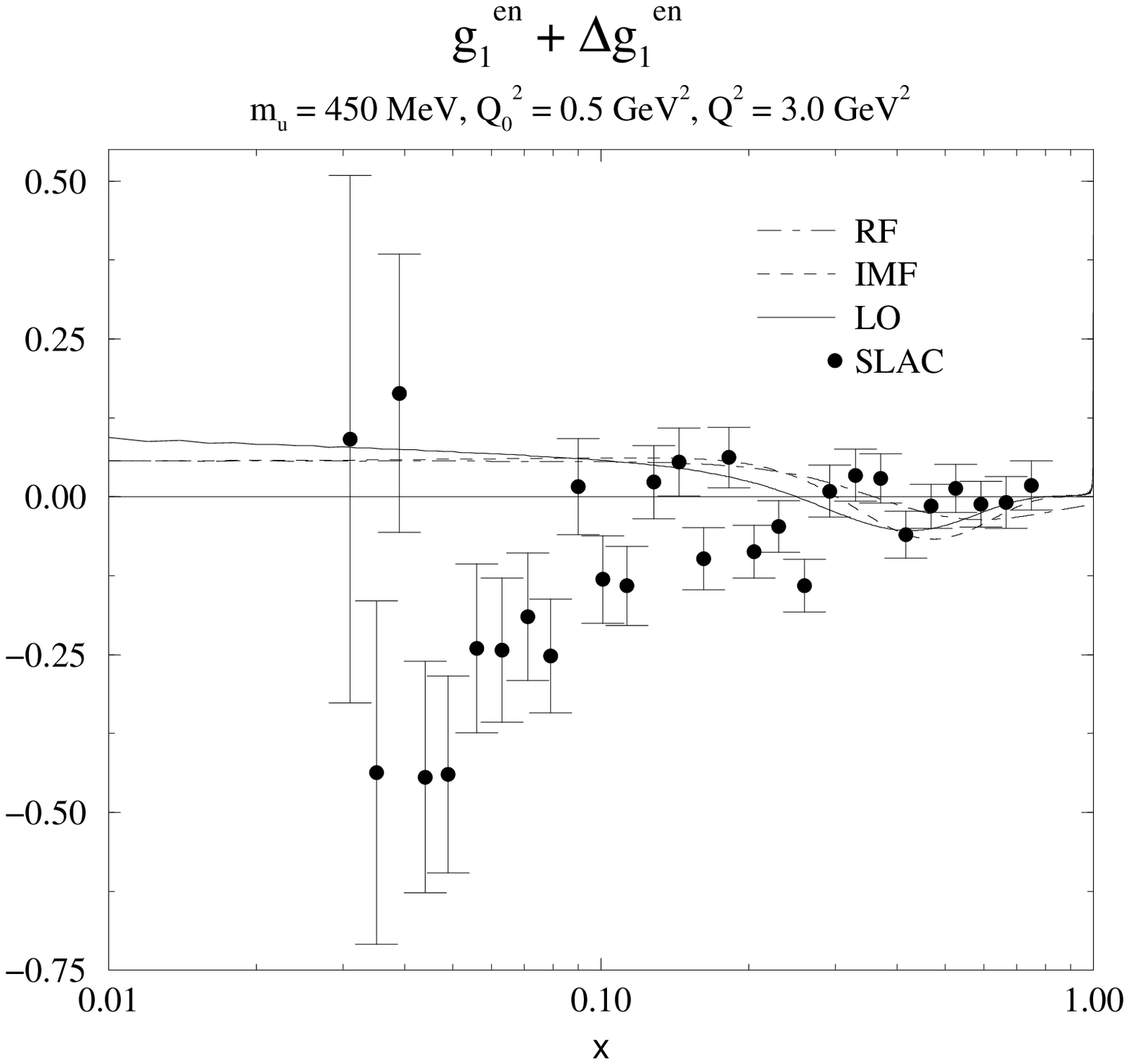,height=6cm,width=7.5cm}}
\end{figure}

As already pointed out in the introduction the polarized structure 
functions are of particular interest. Here we present the results
for both electron--proton as well as electron--neutron scattering
in figures \ref{fig_g1ep} and \ref{fig_g1en}, respectively.
We see that after projection and DGLAP evolution the results
for electron--proton scattering reproduce the gross features 
of the empirical data \cite{SLAC98} for both values of the 
constituent quark mass $m=400{\rm MeV}$ and $450{\rm MeV}$. 
However, at moderate $x$ the predictions slightly overestimate
the data while at larger $x$ they drop off too quickly. 
This might be linked to the boost into the IMF which strongly squeezes 
the structure functions as the comparison between the RF and 
IMF structure functions in figure \ref{fig_g1ep} indicates.
This is especially the case for smaller constituent quark masses
as for those the RF structure functions are already moderately
localized. Other projection procedures \cite{Tr95} may have a 
weaker effect. We would also like to mention that the bilocal
corrections $\triangle g_1(x)$ tend to enhance this pattern.

By taking appropriate matrix elements in eqs (\ref{tg1}) and (\ref{dg1}) 
we straightforwardly obtain the polarized structure functions for 
electron--neutron scattering.
When comparing these results to the data of the E143 experiment
\cite{SLAC98} in figure \ref{fig_g1en} we observe that the model
properly reproduces the small trough in the experimental data at 
$x\approx0.5$. On the other hand our model calculation does not
share the tendency of most of the data to become sizable and negative 
at smaller $x$. To us it seems that the errors on the data are still 
too large to draw very stringent conclusions in this respect.

\begin{figure}[th]
\caption{{\sf \label{fig_g1s}Same as figure \protect\ref{fig_f1s}
for the polarized structure functions. Here both the valence and 
the vacuum part of the inertia parameters enter the expressions
for the angular velocity (\protect\ref{rgen},\protect\ref{omega8}).}}
\vspace{0.5cm}
\centerline{\hspace{0cm}
\epsfig{figure=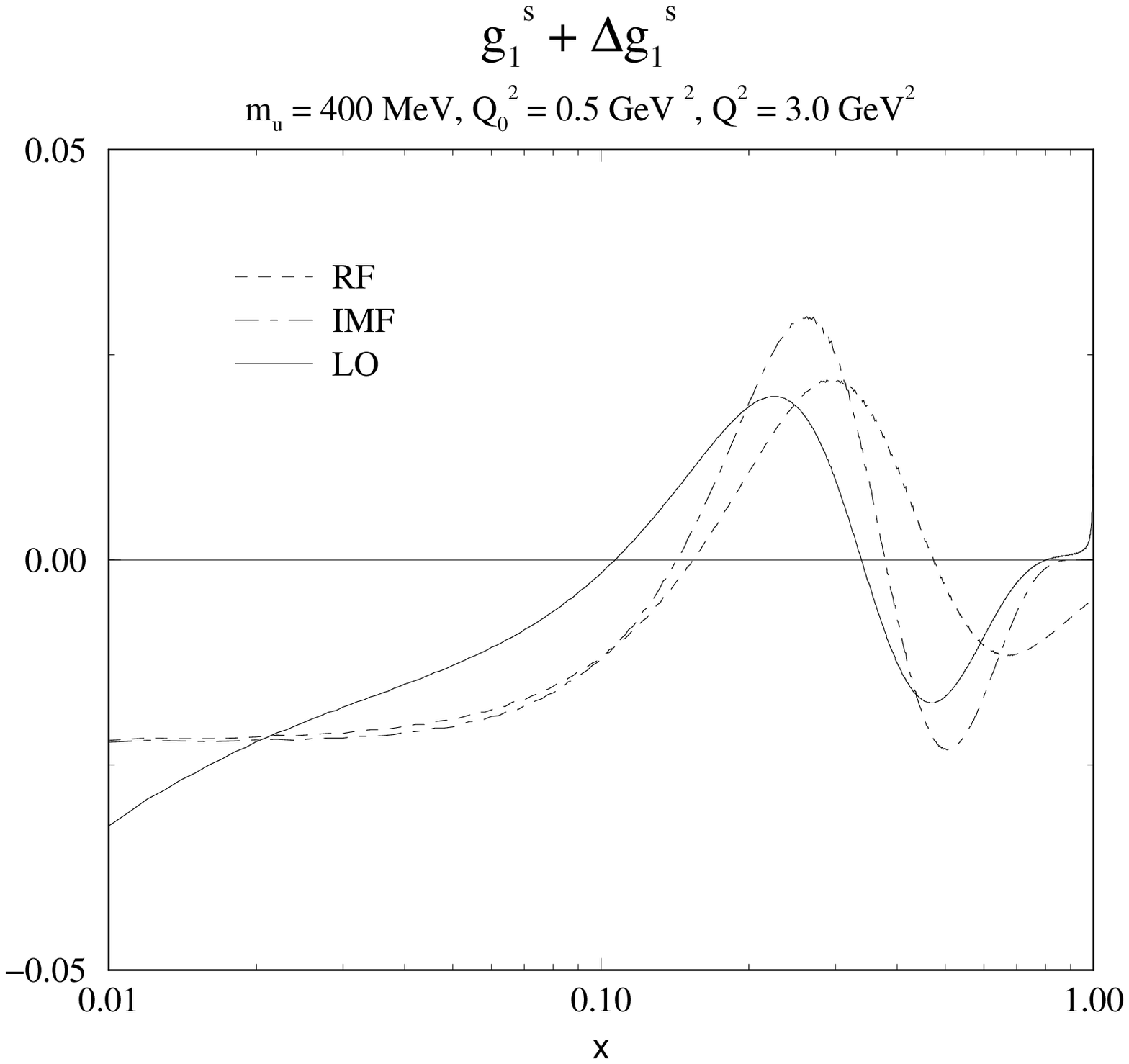,height=6cm,width=7.5cm}
\hspace{1cm}
\epsfig{figure=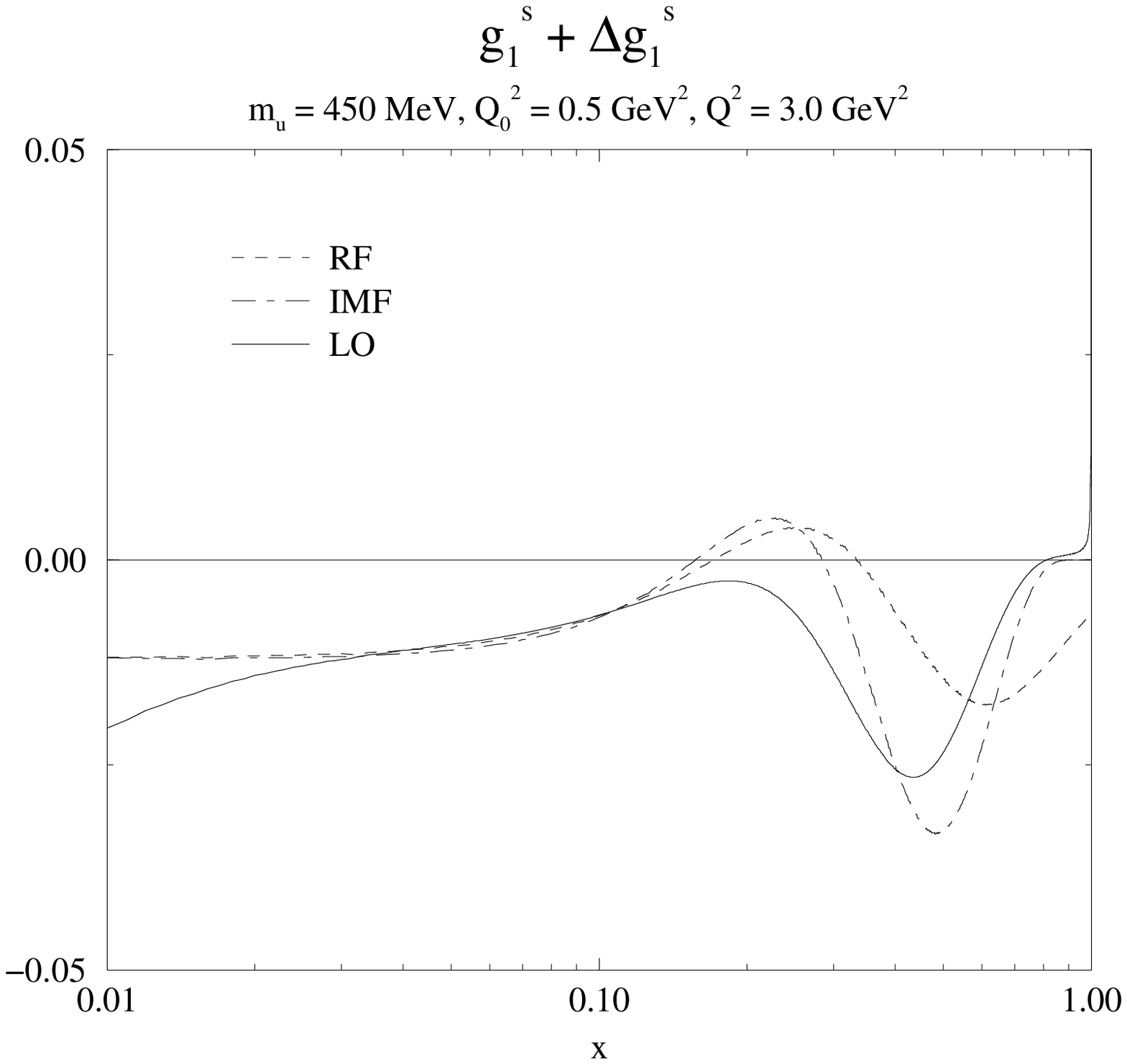,height=6cm,width=7.5cm}}
\end{figure}

Finally we turn to the strange quark contribution of the polarized
structure function. Again this is defined by setting
${\cal O}^{ab} = {\cal O}^{ba} = (1-\sqrt{3} \lambda_8)/3 $. Our 
results are shown in figure \ref{fig_g1s}. These have already
been presented previously \cite{Sch98}, however, with the bilocal 
correction $\triangle g_1^s(x)$, which is sub--leading in $1/N_C$, omitted.
The general shape of $g_1^s(x)+\triangle g_1^s(x)$ is not altered 
by this additional piece. In particular the significant variation
with the constituent quark mass $m$ is maintained. We observe
from figure \ref{fig_g1s} that this structure function has 
the strong tendency to become more negative at larger $x$ with 
increasing $m$. In any event, we should emphasize that the 
magnitude of the strangeness contribution to $g_1(x)$ is quite
small for all permissible parameters and the conclusion of 
a more or less vanishing polarization of the strange quarks
in the nucleon is conceivable. 
\begin{figure}[th]
\caption{{\sf \label{fig_g1s2a3}Comparison with experiment 
\protect\cite{SLAC98} of the two and three flavor model predictions 
for the polarized structure function $g_1(x)$ of electron nucleon 
scattering.}}
\vspace{0.5cm}
\centerline{\hspace{0cm}
\epsfig{figure=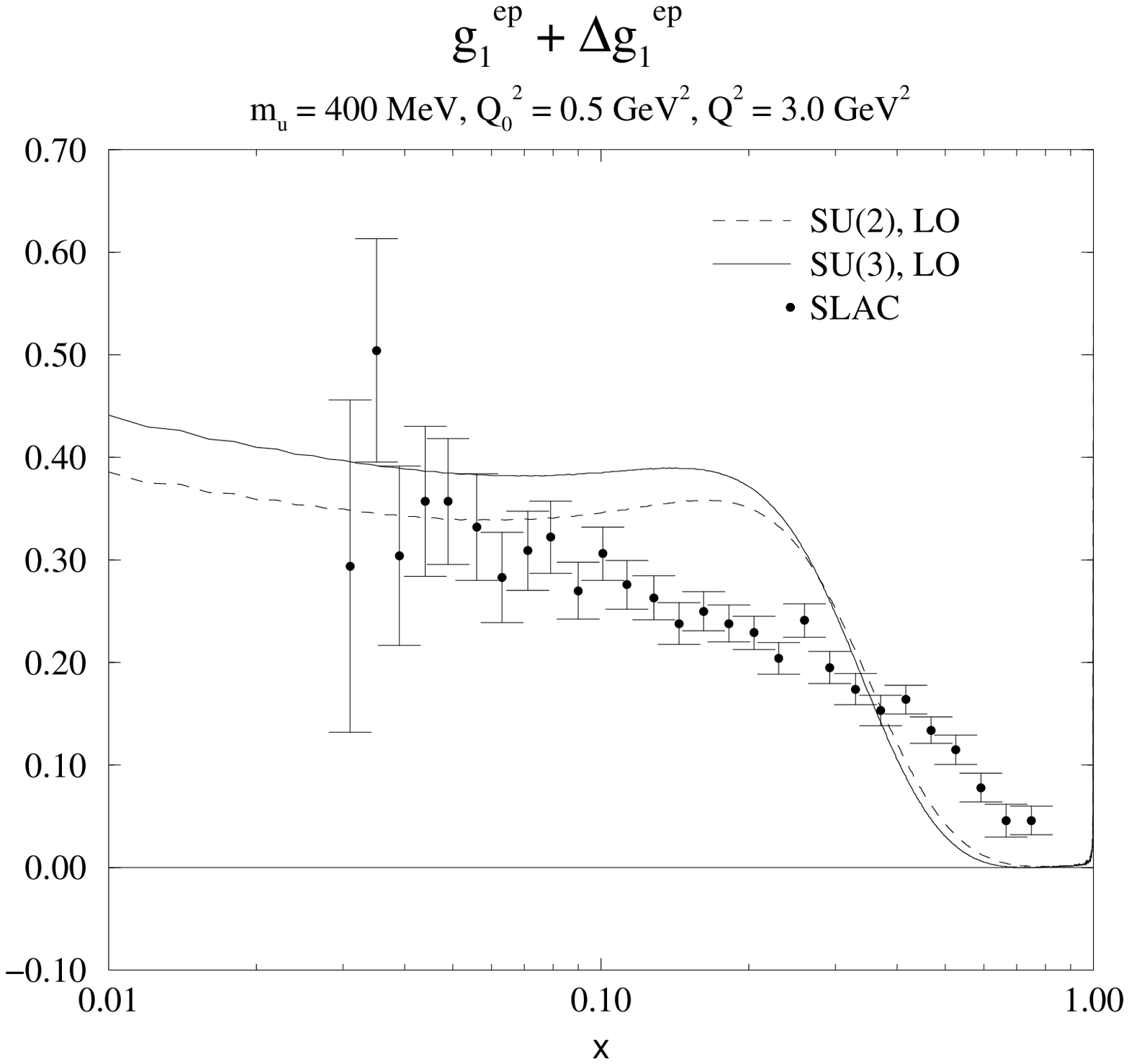,height=6cm,width=7.5cm}
\hspace{1cm}
\epsfig{figure=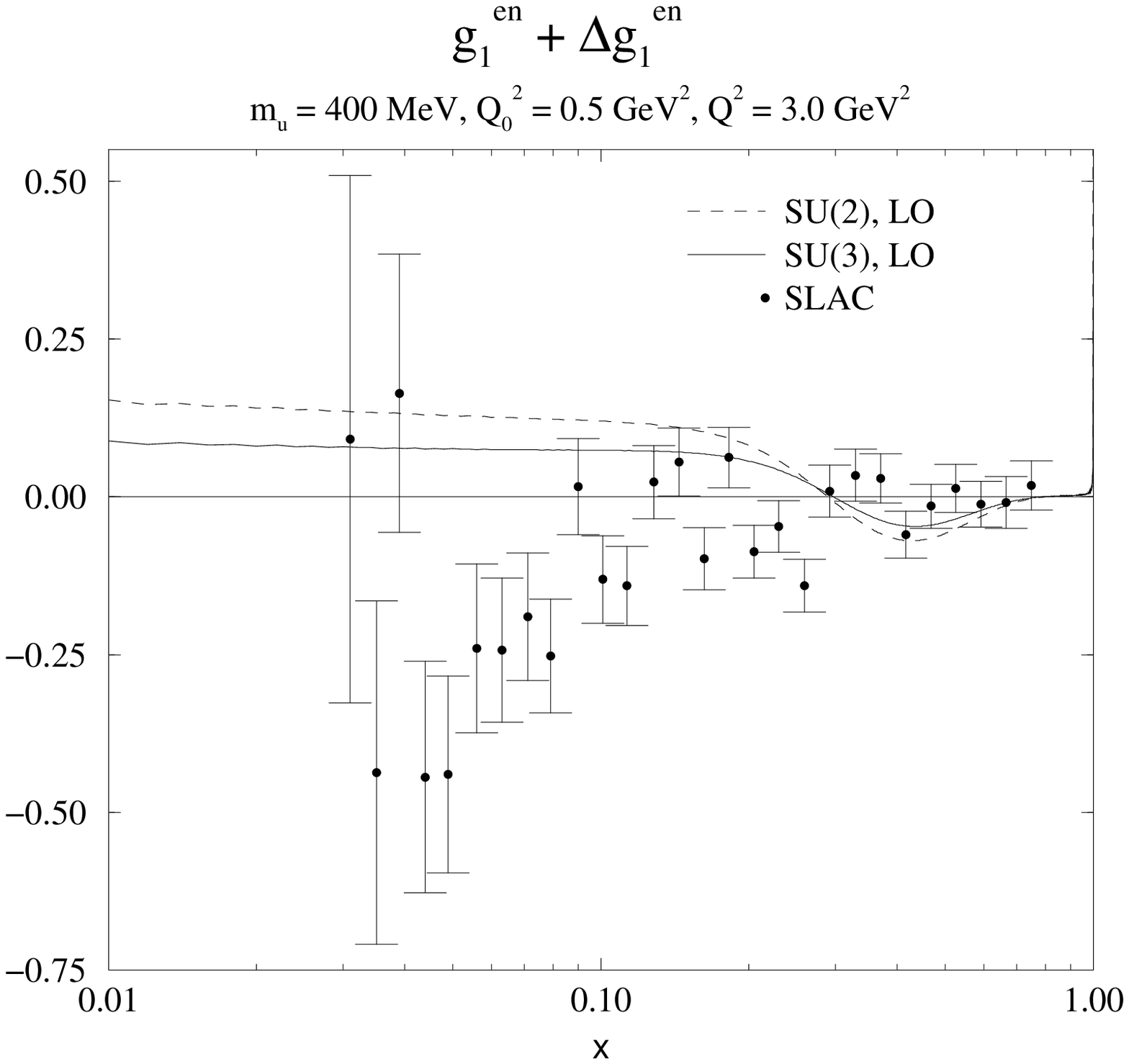,height=6cm,width=7.5cm}}
\end{figure}
The conclusion that strangeness
degrees of freedom yield at most moderate effects can
also be drawn from figure \ref{fig_g1s2a3} where we compare
our present results for the polarized structure functions to
those obtained in the two flavor reduction of the same model.
The isoscalar combination $g_1^{\rm ep}+g_1^{\rm en}$ remains
almost unchanged by the inclusion of strange quarks while there 
is a slight increase for the isovector combination
$g_1^{\rm ep}-g_1^{\rm en}$. The fact that the latter changes at all 
is not only due to the additional terms ({\it cf.} appendix C) 
when extending the model to three flavors but also to the 
change between SU(2) and SU(3) Clebsch--Gordan coefficients  
arising in the calculation of nucleon matrix elements of collective 
operators\footnote{When constructing the exact eigenstates
of the collective Hamiltonian we employ the full inertia
parameters when applying the valence quark approximation
to the structure functions.}. We already communicated this 
result in ref \cite{Sch98} for the case of the polarized 
structure function $g_1(x)$. As can be observed from figure 1 in 
that reference, the use of flavor symmetric collective 
wave--functions enhances the difference between the extrema of 
$g_1(x)$ by almost a factor two. 

The smallness of the strange quark contributions to the nucleon
structure functions is (at least partially) due to fact that 
we employ a nucleon wave--function obtained by exactly 
diagonalizing the collective Hamiltonian. This mitigates the role 
of strange quarks in the nucleon \cite{We96r}. We have checked that
using an SU(3) symmetric wave--function yields larger strange quark
pieces in the nucleon structure functions.

\begin{figure}[ht]
\caption{{\sf \label{fig_gluon}Gluon contributions to the 
unpolarized (left panel) and polarized (right panel) 
structure functions as they arise from the DGLAP 
evolution (\protect\ref{ggs}). Note the different scales 
for the unpolarized and polarized contributions.}}
\vspace{0.5cm}
\centerline{\hspace{0cm}
\epsfig{figure=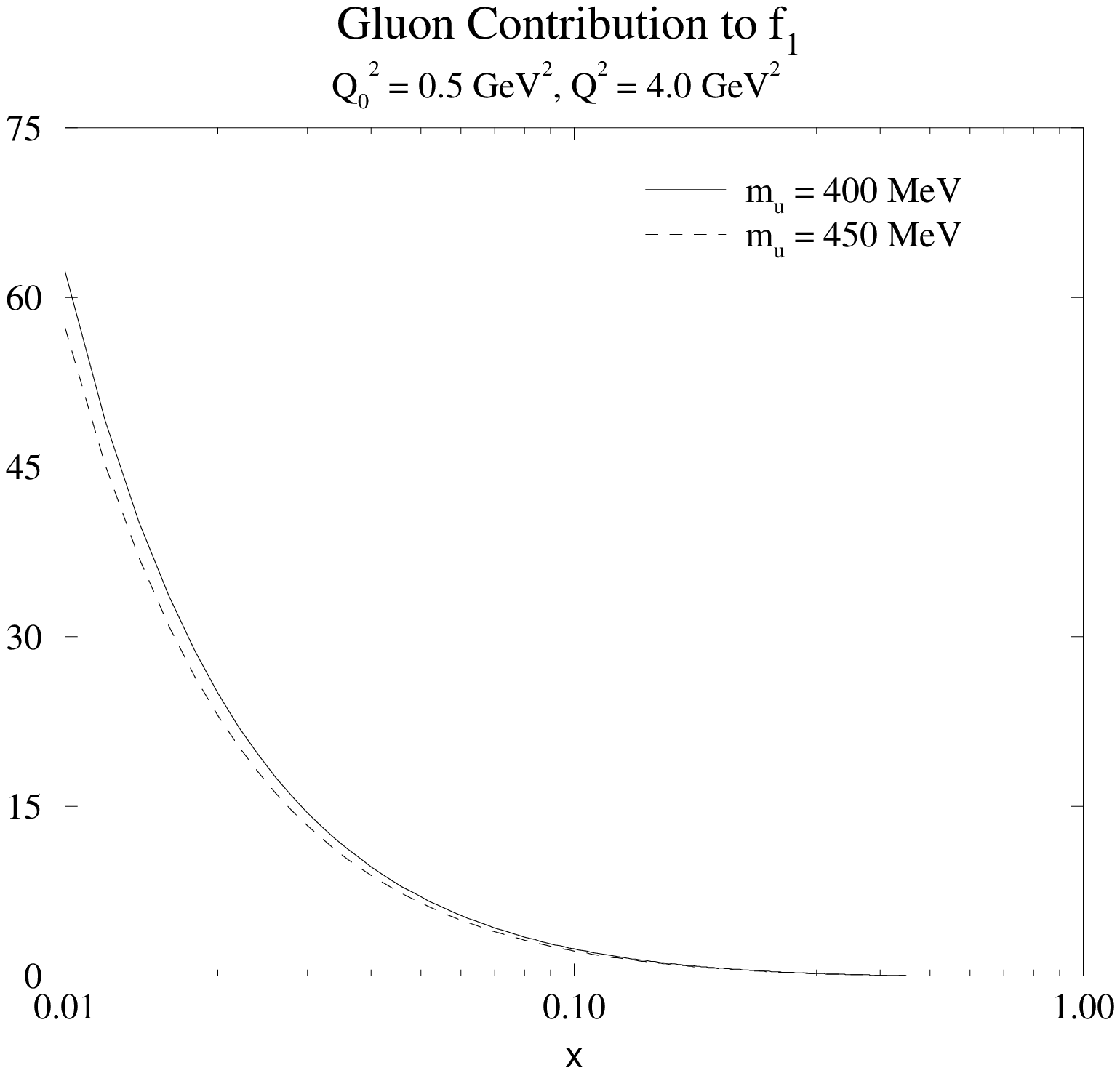,height=6cm,width=7.5cm}
\hspace{1cm}
\epsfig{figure=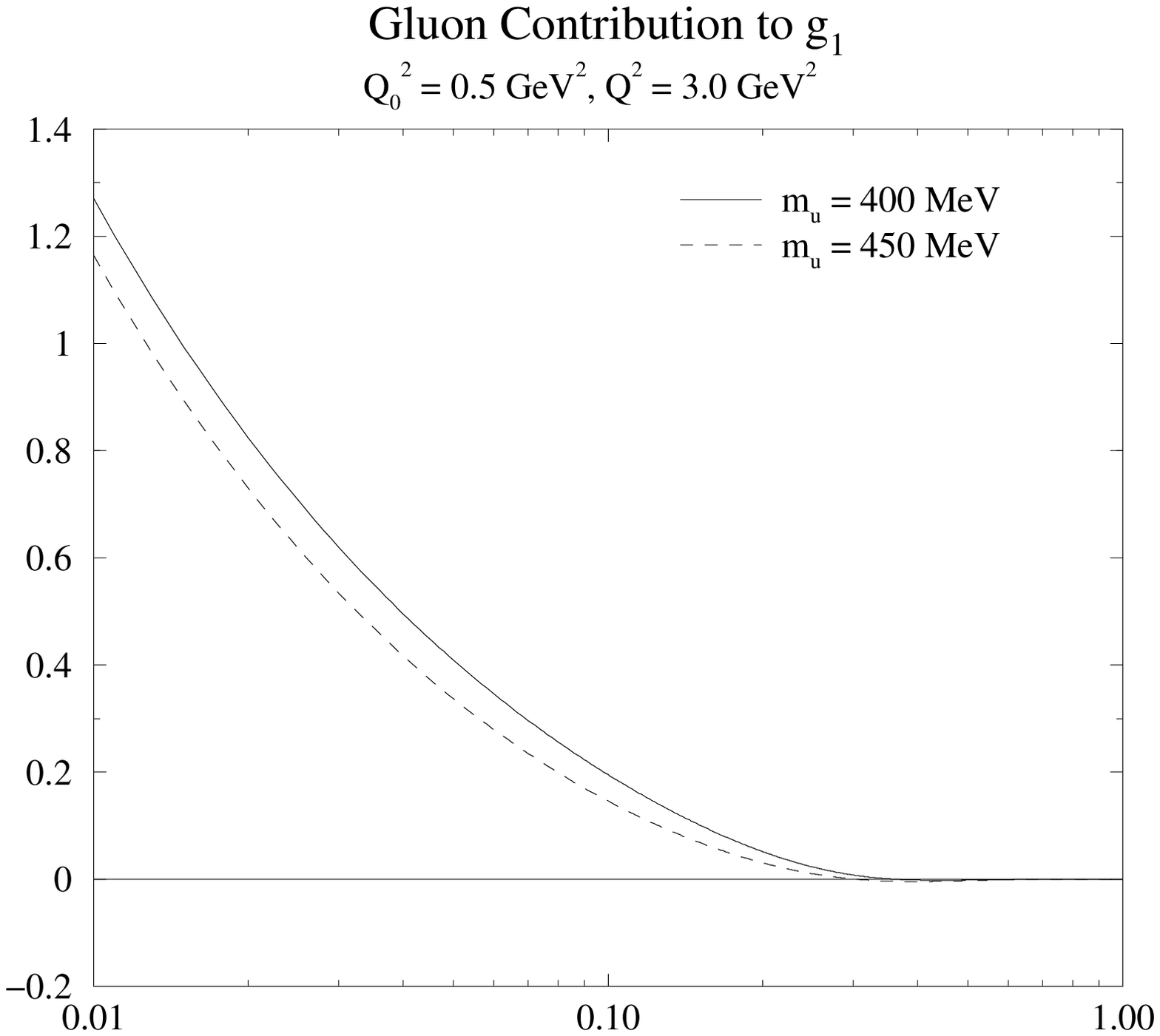,height=6cm,width=7.5cm}}
\end{figure}
{}For completeness we also show the gluon contributions to the
structure functions at the large momentum scale $Q^2$ as they
emerge from the DGLAP evolution (\ref{ggs}) in figure
\ref{fig_gluon}. We recall that as an assumption these 
contributions were put to zero at the model scale $Q^2_0$.
Apparently the gluonic piece in the unpolarized structure function
is almost two orders of magnitudes larger than in the polarized
case. This huge gluonic contribution to the unpolarized structure
function due to the DGLAP evolution is a well known effect. On a 
double logarithmic plot the unpolarized gluon distribution (times
Bjorken $x$) grows approximately linearly as $x\to 0$ for large 
enough $Q^2$ \cite{Gl98}. The dependence of the gluonic pieces on 
the constituent quark mass apparently is negligible.

\bigskip
\leftline{\Large\it 7. Conclusion}
\medskip

We have studied the nucleon structure functions within the three 
flavor Nambu--Jona--Lasinio chiral soliton model which in particular
allowed us to compute the contribution of strange quarks to the
structure functions. 

Like in QCD, in this model the flavor symmetry is explicitly broken by 
different current quark masses. By the mass generation mechanism 
of spontaneous breaking of chiral symmetry this also induces flavor 
symmetry breaking for the constituent quark masses. In the baryon 
sector of the model the flavor symmetry breaking effects are 
explicitly incorporated by using an extension of the Yabu--Ando 
approach to quantize the chiral soliton. This yields a collective 
wave--function for the nucleon with good flavor and spin quantum 
numbers. In addition a projection onto good momentum states is 
performed to restore translational invariance which is broken by
the soliton. Having obtained states of good momentum, the structure
functions computed from a localized object are boosted to the 
infinite momentum frame.

Although for the construction of the chiral soliton the inclusion of 
the Dirac sea is mandatory, in the evaluation of the pertinent matrix 
elements of the hadronic tensor we use the valence quark approximation 
for the nucleon wave--function. This not only avoids the unsolved 
problem of consistently regularizing the vacuum contribution but 
should also be sufficient since in the 
chiral NJL soliton model the static properties of the nucleon are dominated 
or even saturated by their valence quark contribution. In this context 
we should emphasize, however, that in our model the valence quarks are 
bound constituent quarks and thus differ essentially from a valence 
(or sea) quark in the parton model. To minimize the ambiguities when 
identifying model and QCD degrees of freedom at the low energy scale
where our model is supposed to approximate QCD we only use the leading 
order DGLAP formalism to evolve the predicted structure functions 
to the energy scale encountered in the corresponding experiments.

In the numeric evaluation we adjust the parameters of the model to 
well--known mesonic properties leaving only the constituent up quark
mass as a free parameter. Note that this also determines the constituent
strange quark mass. The Gottfried sum rule is fairly well reproduced for 
a constituent quark mass of $m=400{\rm MeV}$ but somewhat underestimated 
for $m=450{\rm MeV}$.  This underestimation partially reflects the fact 
that for large constituent quark masses the contribution of the distorted
vacuum becomes non--negligible (In this context one should note that the 
constituent quark mass represents the coupling constant between the 
quarks and the chiral field). The predicted unpolarized structure functions 
which enter the Gottfried sum rule overestimated the experimental data at 
large and moderate $x$ but underestimate them at small $x$. A better 
agreement could probably be obtained by lowering the reference scale of 
the model in the DGLAP evolution. However, then one would have to go 
beyond the one leading order splitting functions.

A main focus of the present paper has been the contribution of the 
strange quark to the unpolarized and polarized structure functions. 
The strange quark part to the unpolarized structure functions 
$f_{1}^{s} (x)$ exhibits a pronounced peak at $x = 0.5$. The shift 
from the na\"\i vely expected peak at $x=1/3$ can be attributed to the 
larger mass of the strange quark. The DGLAP evolution leads only to a
moderate smearing  of the peak. At lower $x$ the strangeness contribution 
to the unpolarized structure functions is negative which is probably 
due to the valence quark approximation which ignores certain 
(in the vacuum) disconnected diagrams which in the presence of the 
chiral soliton field become connected. In general, however, the 
{\underline{regularized}} vacuum contributions need not restore
positivity. The DGLAP evolution substantially 
increases the strangeness contribution at smaller $x$.

After projection and DGLAP evolution the calculated polarized structure 
function for deep inelastic electron--proton scattering reproduces the 
gross features of the experimental data, although our results slightly 
overestimate the data at moderate $x$ while they drop off somewhat too 
quickly at larger $x$. For electron--neutron scattering our calculated 
polarized structure function properly reproduces the small trough in
the experimental data at $x \approx 0.5$ but fail to become sizable 
enough at lower $x$. Unfortunately the error bars on the data seem
to be too large to allow for very stringent conclusions. Again,
choosing a smaller reference scale of the model might improve on 
these deviations from the data. Our numerical 
results show that the strange quark contribution of the polarized 
structure functions is quite small and in fact conceivable with a
vanishing polarization of the strange quarks in the nucleon. The 
comparison of the results from the two and three flavor chiral soliton
model leads us to the conclusion that strange degrees of freedom yield 
at most moderate effects to the structure functions of the nucleon.

\bigskip
\leftline{\Large\it Acknowledgments}
\medskip
Helpful discussions with Leonard Gamberg are gratefully acknowledged.

\bigskip
\leftline{\Large\it A. Basis wave--functions}
\medskip
Although the basis states to diagonalize the quark Dirac Hamiltonian 
(\ref{hamil}) in the background of the static soliton may be traced 
from the literature ({\it cf.} ref \cite{Ka84}) we repeat them here for
completeness and use in the proceeding appendices. The eigenfunctions 
of this static Hamiltonian are eigenfunctions of the grand spin operator
\begin{equation}
\vec{G} = \vec{j} + \frac{\vec{\tau}}{2} = 
\vec{l} + \frac{\vec{\sigma}}{2} +
\frac{\vec{\tau}}{2}  
\end{equation}
where ${\vec\tau}/2$ denotes the isospin operator in the fundamental
representation. It is easily verified that $\vec G$ commutes 
with single particle Hamiltonian (\ref{hamil}).
Additionally, the wave--functions are eigenfunctions of the
parity operator. To describe the coupling 
of orbital angular momentum $l$, spin $j=l\pm1/2$ and
isospin $\tau =1/2$ to the total grand spin $G$ with projection
quantum number $M$ we adopt the notation
\begin{equation}
 {\cal Y}^{GM}_{lj}(\hat{r}) = \langle \hat{r} | ljGM \rangle
\qquad {\rm with} \qquad
| ljGM \rangle = \sum_{j_3,\tau_3} C^{GM}_{jj_3 \frac{1}{2}\tau_3} |ljj_3
\rangle | \frac{1}{2}\tau_3 \rangle \, .
\end{equation}
The coordinate space wave--functions for the (non--strange)
up and down quarks are given by
\begin{eqnarray}
\Psi_{\mu}^{(G,+)}(\vec r) = \left(\begin{array}{lr} 
i g_{\mu}^{(G,+;1)}(r) &
{\cal Y}^{GM}_{G G+1/2}(\hat r) \\ f_{\mu}^{(G,+;1)}(r)& {\cal Y}^{GM}_{G+1
G+1/2}(\hat r) \end{array} \right) +  \left( \begin{array}{lr} i
g_{\mu}^{(G,+;2)}(r) &  {\cal Y}^{GM}_{G G-1/2}(\hat r) \\
-f_{\mu}^{(G,+;2)}(r)& {\cal Y}^{GM}_{G-1 G-1/2}(\hat r) 
\end{array} \right) 
\end{eqnarray} 
\begin{eqnarray}
\Psi_{\mu}^{(G,-)}(\vec r) = \left( \begin{array}{lr} 
i g_{\mu}^{(G,-;1)}(r) &
{\cal Y}^{GM}_{G+1 G+1/2}(\hat r) \\ -f_{\mu}^{(G,-;1)}(r) & 
{\cal Y}^{GM}_{G G+1/2}(\hat r) \end{array} \right) +  
\left( \begin{array}{lr} i g_{\mu}^{(G,-;2)}(r) &  
{\cal Y}^{GM}_{G-1 G-1/2}(\hat r) \\
f_{\mu}^{(G,-;2)}(r) & {\cal Y}^{GM}_{G G-1/2}(\hat r) \end{array} \right)
\end{eqnarray} 
The superscript $\pm$ refers to the intrinsic parity $\Pi_{\rm int}$
which is defined via total parity and the grand spin 
$\Pi=\Pi_{\rm int}(-1)^G$.
To discretize the quark--orbits we demand that the upper components of 
these spinors to vanish at the boundary of a spherical cavity with 
radius $D$\footnote{Obviously that radius is chosen to be a 
multiple of the typical soliton extension.}.
It should be noted that the classical part of the valence quark 
wave--function dwells in the $G=0$ channel with only $j=1/2$ being 
permitted:
\begin{eqnarray}
\Psi_{\rm V}(\vec r) = 
\left(
\begin{array}{lr} i g_{\rm V}(r) & {\cal Y}^{00}_{0, 1/2}(\hat r)\\ 
f_{\rm V} (r) & {\cal Y}^{00}_{1, 1/2}(\hat r) \end{array} \right)\, .
\end{eqnarray}
As a shorthand notation  to deal with the symmetry breaking corrections 
we introduce here the so--called {\it chirally rotated}
radial wave--functions
\begin{eqnarray}
\label{chged}
g^{\Theta_c} = g \cos{\frac{\Theta_c}{2}} 
+ f \sin{\frac{\Theta_c}{2}} \quad {\rm and}\quad
f^{\Theta_c} = f \cos{\frac{\Theta_c}{2}}
-g \sin{\frac{\Theta_c}{2}} 
\end{eqnarray}
Here $\Theta_c$ denotes the chiral angle which minimizes the 
classical energy (\ref{efunct}).
Since the classical soliton is embedded in the isospin subspace, 
the strange quarks wave--functions are those of a free Dirac field
in a spherical cavity. We choose spinors with good angular momentum 
({\it cf.} \cite{Al94a}):
\begin{eqnarray*}
s_{nl}^1(\vec{r}) & = & \langle \vec{r} |1,n,j=l + \frac{1}{2}, m \rangle =  
{\cal N}_{nl} \left( \begin{array}{rl} i \bar{w}_{nl}^{+}j_l(k_{nl}r)
&\langle \vec{r} | l l+\frac{1}{2} m \rangle \\  
\bar{w}_{nl}^{-}j_{l+1}(k_{nl}r) &\langle\vec{r}|l+1 l+\frac{1}{2}m\rangle
  \end{array} \right) \\ 
s_{nl}^2(\vec{r}) & = & \langle \vec{r} |2,n,j=l - \frac{1}{2}, m \rangle  
=  {\cal N}_{nl} 
\left( \begin{array}{rl} i \bar{w}_{nl}^{+}j_l(k_{nl}r) &\langle \vec{r} | l
    l-\frac{1}{2}  m \rangle \\  
   -\bar{w}_{nl}^{-}j_{l-1}(k_{nl}r) &\langle \vec{r} | l-1 l-\frac{1}{2} m
   \rangle 
  \end{array} \right) \\  
\end{eqnarray*}
with
\begin{eqnarray*}
E_{nl} & = & \pm \sqrt{k^2_{nl}+m_s^2} \hspace*{4em} {\cal N}_{nl}  
= \frac{1}{D^{3/2}|j_{l+1}(k_{nl}D)|} \\
\bar{w}^{+}_{nl}& = & \sqrt{1+m_s/E_{nl}}, \hspace*{3em} \bar{w}^{-}_{nl} =
{\rm sgn}(E_{nl})\sqrt{1-m_s/E_{nl}} \, .
\end{eqnarray*}

Although the diagonalization of the static quark Dirac Hamiltonian 
operator is most conveniently performed in position space we will need 
the wave functions in momentum space for the computations of the 
structure functions. The grand spin structure will be preserved 
under the Fourier transformation. Generally we denote by $\tilde R(p)$ 
the Fourier transform of the radial function function $R(r)$ keeping 
in mind that it implements a spherical Bessel function of the
order  
of the associated orbital angular momentum, {\it e.g.} 
$\tilde g_\mu^{(G,+,1)}(p)=\int_0^D r^2 dr g_\mu^{(G,+,1)}(r) j_G(pr)$.
For the strange quarks the situation is a bit more involved and 
we find it suitable to define $\tilde{s}_{nl}(k)$ as the Fourier 
transform involving $j_{l}(k_{nl}r)$ while $\tilde{t}^1_{nl}(k)$ 
is obtained from integrating over $j_{l+1}(k_{nl}r)$ and 
$\tilde{t}^2_{nl}(k)$ contains $j_{l-1}(k_{nl}r)$. 

\newpage
\bigskip
\leftline{\Large\it B.
Valence quark wave--function:}

\smallskip
\leftline{\Large\it ~~~~cranking and symmetry breaking}

Here we make explicit the corrections to the classical valence
quark wave--function which are due to both the collective rotation
$h_{\rm rot}$ and the flavor symmetry breaking part $h_{\rm SB}$
of the perturbation in eq (\ref{hmod}) (see also eq (\ref{valwfct})).
\begin{eqnarray*}
&&\hspace{-0.5cm}\sum_{\nu \neq \rm V}  \Psi_{\rm V}\,
\frac{\langle \nu | h_{\rm rot}+h_{\rm SB} |{\rm V} \rangle}
{\epsilon_{\rm V} - \epsilon_{\nu}} =\\*
&&\sum_{\mu \neq {\rm V}} \Psi_{\mu} \Bigg\{ \delta_{G1} 
\Bigg(
Q_{\mu} \left[\frac{\delta_{M1}}{\sqrt{2}} (\Omega_1 - i \Omega_2) -
\frac{\delta_{M-1}}{\sqrt{2}} (\Omega_1 + i \Omega_2) -
\delta_{M0} \Omega_3\right]  \\
&& \hspace*{5em} + P_{\mu} \left[\frac{\delta_{M1}}{\sqrt{2}}
(D_{81} - i D_{82}) 
- \frac{\delta_{M-1}}{\sqrt{2}} (D_{81} +iD_{82}) 
- \delta_{M0} D_{83}\right]\Bigg) 
+ \delta_{G0} \Gamma_{\mu} (D_{88}-1) \\
&& \hspace*{2em}-\frac{{\cal N}_{nl}/\sqrt{2}}
{\epsilon_{\rm V}-\epsilon_{nl}} \delta_{l0}\delta_{j\frac{1}{2}}
\Bigg( \frac{1}{2} \left[\delta_{m_j \frac{-1}{2}}(\Omega_4 + i \Omega_5)
- \delta_{m_j \frac{1}{2}} (\Omega_6 + i \Omega_7)\right] 
\left[{\bar w}_{nl}^+\tilde{g}_{\rm V}(k_{nl})+{\bar w}_{nl}^-
  \tilde{f}_{\rm V}(k_{nl})\right] \\
&&\hspace*{2.7em}+\frac{m-m_s}{\sqrt{3}} \left[\delta_{m_j \frac{-1}{2}}
(D_{84} + i D_{85}) 
-\delta_{m_j \frac{1}{2}} (D_{86} + i D_{87})\right]
\left[w_{nl}^+ \tilde{g}^{\Theta_c}_{\rm V}(k_{nl}) +
 w_{nl}^- \tilde{f}^{\Theta_c}_{\rm V}(k_{nl})\right] \Bigg) 
\Bigg\} \, .
\end{eqnarray*}
The summation index $\mu$ runs over $\mu,G$ and $M=-G,..,+G$ when 
coupling to a non--strange spinor and over $n,j=l\pm1/2$ and 
$m_j=-j,..+j$ in the other case. We have used the following 
abbreviations for the overlap integrals:
\begin{eqnarray*}
Q_{\mu} & = & \frac{1}{\epsilon_{\rm V}-\epsilon_{\mu}} \int
drr^2\left(g_{\rm V}g^{(2,-)}_{\mu}+f_{\rm V}f^{(2,-)}_{\mu}\right)\\
P_{\mu}&=&\frac{m-m_s}{\sqrt{3}(\epsilon_{\rm V}-\epsilon_{\mu})} 
\int drr^2\left(g_{\rm V}^{\Theta_c} g^{(2,-) \Theta_c}_{\mu}
-f_{\rm V}^{\Theta_c} f^{(2,-)\Theta_c}_{\mu}\right)\\
\Gamma_{\mu} & = & \frac{m-m_s}{3(\epsilon_{\rm _V}-\epsilon_{\mu})}
\int drr^2\left(g_{\rm V}^{\Theta_c} g^{(1,+) \Theta_c}_{\mu} 
-f_{\rm V}^{\Theta_c} f^{(1,+) \Theta_c}_{\mu}\right) 
\end{eqnarray*}
while the momenta $k_{nl}$ are the roots of 
$j_l(k_{nl}D)$.

\bigskip
\leftline{\Large\it C. Expressions for the 
structure functions}
\medskip
In this appendix we present the expressions entering
the structure function calculation using the wave--functions
of the preceding appendices.

\medskip
\leftline{\large\it C.1 Explicit expressions for f$_1$}
\medskip
In this part we give the explicit expression for the unpolarized
structure function $f_1$ in terms of the radial wave--functions that
have been described in Appendix A. Again, $g^{\Theta_c}$ and
$f^{\Theta_c}$ denote the chirally rotated wave functions.
\begin{eqnarray*}
f_1 & = & \delta_0' f_{1,-}^0 + \delta_3' f_{1,-}^3 + \delta_8' f_{1,-}^8 -
  \delta_0 f_{1,+}^0 - \delta_3 f_{1,+}^3 - \delta_8 f_{1,+}^8 \ \ {\rm
  with} 
\\  
f_{1,\pm}^0 & = & N_C \frac{M_N}{\pi} \langle N | 
\int_{M_N |x_{\pm}|}^{\infty} p dp d\phi\, 
\tilde{\psi}_{\rm V}^{\dagger} (\vec{p}_{\pm}) (1 \pm \alpha_3)
\tilde{\psi}_{\rm V} (\vec{p}_{\pm}) | N \rangle \\
f_{1,\pm}^l & = & N_C \frac{M_N}{\pi} \langle N | 
\sum _{a=1}^8D_{la} 
\int_{M_N |x_{\pm}|}^{\infty} p dp d\phi\,
\tilde{\psi}_{\rm V}^{\dagger}(\vec{p}_{\pm})(1\pm\alpha_3)\lambda_a
\tilde{\psi}_{\rm V} (\vec{p}_{\pm}) | N \rangle \, ,
\qquad {\rm for}\quad l=3,8\, .
\end{eqnarray*}

\begin{eqnarray*}
f_{1,\pm}^0 & = & N_C \frac{M_N}{\pi} 
\int_{M_N |x_{\pm}|}^{\infty}  p dp \left(
\tilde{g}_{\rm V}^2 \mp 
2\tilde{f}_{\rm V} \tilde{g}_{\rm V}
\cos{\Theta^{\pm}} +  \tilde{f}_{\rm V}^2 \right) \\*
& & \hspace*{-1em} +  \sum_{\mu \ne {\rm V}} 
\Gamma_{\mu} \left[\tilde{g}_{\rm V} 
\tilde{g}^{(1,+)}_{\mu} + \tilde{f}_{\rm V} \tilde{f}^{(1,+)}_{\mu} \mp
\cos{\Theta^{\pm}}\left(\tilde{g}_{\rm V} \tilde{f}^{(1,+)}_{\mu} 
+\tilde{f}_{\rm V} \tilde{g}^{(1,+)}_{\mu}\right)\right] 
\langle N |(D_{88}-1) | N \rangle 
\\
f_{1,\pm}^l & = & N_C \frac{M_N}{\pi} 
\int_{M_N |x_{\pm}|}^{\infty} p dp \Bigg\{
\left(\tilde{g}_{\rm V}^2 \mp 2 \tilde{f}_{\rm V} \tilde{g}_{\rm V}
\cos{\Theta^{\pm}} +  \tilde{f}_{\rm V}^2  \right)
\frac{\langle N | D_{l8} | N \rangle}{2 \sqrt{3}} \\*
&&+\sum_{\mu \ne {\rm V}}\Bigg(\delta_{G0}\Gamma_\mu
\left[\tilde{g}_{\rm V}  \tilde{g}^{(1,+)}_{\mu}
+\tilde{f}_{\rm V}\tilde{f}^{(1,+)}_{\mu} 
\mp\cos{\Theta^{\pm}} \left(\tilde{g}_{\rm V}\tilde{f}^{(1,+)}_{\mu}
+\tilde{f}_{\rm V} \tilde{g}^{(1,+)}_{\mu}\right)\right]
\langle N |D_{l8} (D_{88}-1) | N \rangle  \\
&& + \frac{1}{2}\delta_{G1} 
\left[Q_{\mu}\left(B_{\mu} \mp 2 N_{\mu} \cos{\Theta^{\pm}}\right)
 \langle N | D_{li} \Omega_i | N \rangle 
+ 2 P_{\mu} \left(B_{\mu} \mp 2  N_{\mu} \cos{\Theta^{\pm}}\right)
 \langle N | D_{li} D_{8i} | N \rangle \right] 
\\ & & 
+\frac{1}{4} \delta_{l_\mu 0} 
 \frac{{\cal N}^2_{n0}}{\epsilon_{\rm V}-\epsilon_{\mu}} 
 \bigg[\bigg\{\left(
 \bar{w}_{n0}^{+2}\tilde{g}_{\rm V}(k_{n0})\tilde{s}_{n0}
 +\bar{w}_{n0}^{+}\bar{w}_{n0}^-
  \tilde{f}_{\rm V}(k_{n0})\tilde{s}_{n0}\right)
 \left(\tilde{g}_{\rm V}\pm\tilde{f}_{\rm V} \cos\Theta^{\pm}\right) 
\\ & & \hspace*{2em}
+\left(\bar{w}_{n0}^{+} \bar{w}_{n0}^{-} \tilde{g}_{\rm V}(k_{n0})
 \tilde{t}_{n0}^{1}+\bar{w}_{n0}^{-2} \tilde{f}_{\rm V}(k_{n0})
 \tilde{t}_{n0}^{1}\right)\left(\tilde{f}_{\rm V}\pm\tilde{g}_{\rm V}
 \cos\Theta^{\pm}\right)\bigg\}
\langle N|D_{l\alpha}\Omega_\alpha|N\rangle\\  
&&\hspace*{1em}
+\bigg\{\left(\bar{w}_{n0}^{+2}\tilde{g}_{\rm V}^{\Theta_c}(k_{n0}) 
 \tilde{s}_{n0}+\bar{w}_{n0}^{+}\bar{w}_{n0}^{-}
 \tilde{f}_{\rm V}^{ \Theta_c}(k_{n0})\tilde{s}_{n0}\right)
 \left(\tilde{g}_{\rm V} \pm \tilde{f}_{\rm V} \cos\Theta^{\pm}\right) 
 \\ & & \hspace*{2em}
+\left(\bar{w}_{n0}^{+}\bar{w}_{n0}^{-}\tilde{g}_{\rm V}^{\Theta_c}(k_{n0})
 \tilde{t}_{n0}^{1}+\bar{w}_{n0}^{-2}\tilde{f}_{\rm V}^{\Theta_c}(k_{n0})
 \tilde{t}_{n0}^{1}\right)\left(\tilde{f}_{\rm V} \pm 
 \tilde{g}_{\rm V}\cos\Theta^{\pm}\right)\bigg\} \\
 & & \hspace*{7em}
\times\frac{2}{\sqrt{3}}\left(m-m_s\right) 
\langle N | D_{l\alpha} D_{8\alpha} | N \rangle \bigg] \Bigg) 
\Bigg\} \\  
\end{eqnarray*}
for $l=3,8$ with the summation conventions $i=1,2,3$ and 
$\alpha=4,\ldots,7$. In case not explicitly stated, the integration 
variable $p$ is the argument of $\tilde{g}_V$ and $\tilde{f}_V$.
We have furthermore used the following shorthand notation:
\begin{eqnarray*}
B_{\mu} & = &\tilde{g}_{\rm V} \tilde{g}^{(2,-)}_{\mu} + 
\tilde{f}_{\rm V} \tilde{f}^{(2,-)}_{\mu} \\
N_{\mu} & = &\tilde{g}_{\rm V} \tilde{f}^{(2,-)}_{\mu} + 
\tilde{f}_{\rm V} \tilde{g}^{(2,-)}_{\mu}
\end{eqnarray*} 
in addition to the quantities $Q_{\mu}$, $P_{\mu}$ and 
$\Gamma_{\mu}$ which were already introduced in appendix B.

\medskip
\leftline{\large\it C.2 Explicit expressions for $\Delta$f$_1$}
\medskip

As usual we split up the structure functions according to 
the flavor content and the two contributions from the forward 
and backward intermediate quark. In the case of $\Delta f_1$ 
this yields
\begin{eqnarray*}
\Delta f_1 = \delta_{0} \Delta f_{1,+}^0 +  \delta_{3} \Delta f_{1,+}^3 +
\delta_{8} \Delta f_{1,+}^8 + \delta_{0}' \Delta f_{1,-}^0 +
\delta_{3}' \Delta f_{1,-}^3 + \delta_{8}' \Delta f_{1,-}^8    
\end{eqnarray*}
with 
$$
\Delta f_{1,\pm}^0= -\frac{N_C}{2 \pi} 
\frac{\partial}{\partial x} \int_{M_N |x_{\pm}|}^{\infty} p dp\,
\left(\tilde{f}_{\rm V}(p)^2 + \tilde{g}_{\rm V}(p)^2 \mp
2\tilde{f}_{\rm V}(p)\tilde{g}_{\rm V}(p)\cos{\Theta^{\pm}}\right)
 \, \Bigg[ \frac{1}{8 \alpha^2} - \frac{1}{12 \beta^2} \Bigg]
$$
while for $l=3,8$ we obtain
\begin{eqnarray*}
\Delta f_{1,\pm}^l & = & -\frac{N_C}{2 \pi} 
\frac{\partial}{\partial x} \int_{M_N |x_{\pm}|}^{\infty} p dp\,
\left(\tilde{f}_{\rm V}(p)^2 + \tilde{g}_{\rm V}(p)^2 \mp
2\tilde{f}_{\rm V}(p)\tilde{g}_{\rm V}(p)\cos{\Theta^{\pm}}\right)
 \\* & &\hspace{1em}\Bigg[-\frac{1}{2\alpha^2}
 \langle N|D_{li}(R_i+\alpha_1 D_{8i})|N\rangle 
 -\frac{1}{4\beta^2}\langle N|D_{l\alpha}(R_{\alpha}
    +\beta_1D_{8\alpha})|N\rangle 
 \\* & & \hspace*{3em}+\frac{1}{\sqrt{3}}\left(\frac{1}{8 \alpha^2} 
 -\frac{1}{12\beta^2}\right)\langle N|D_{l8}|N\rangle
    \Bigg]\, .
\end{eqnarray*}

\medskip
\leftline{\large\it C.3 Explicit expressions for g$_1$}
\medskip
In this appendix we give the explicit expression for the polarized structure
function $g_1$ in terms of the radial wave functions that were described in
Appendix A. As previously $g^{\Theta_c}$ and $f^{\Theta_c}$ denote the 
chirally rotated wave functions.
\begin{eqnarray*}
g_1 & = & \delta_0 g_{1,+}^0 + \delta_3 g_{1,+}^3 + \delta_8 g_{1,+}^8
- \delta_0' g_{1,-}^0 - \delta_3' g_{1,-}^3 - \delta_8' g_{1,-}^8 \ \ 
{\rm with}
\\  
g_{1,\pm}^0 & = & N_C \frac{M_N}{\pi} \langle N | \int_{M_N
  |x_{\pm}|}^{\infty} p dp d\phi\, 
\tilde{\psi}_{\rm V}^{\dagger} (\vec{p}_{\pm}) 
(1\pm\alpha_3)\gamma^5\tilde{\psi}_{\rm V}(\vec{p}_{\pm})|N\rangle \\
g_{1,\pm}^l & = & N_C\frac{M_N}{\pi}\langle N | 
\sum_{a=1}^8 D_{la} \int_{M_N |x_{\pm}|}^{\infty} p dp d\phi\,
\tilde{\psi}_{\rm V}^{\dagger} (\vec{p}_{\pm}) 
(1\pm\alpha_3)\gamma^5\lambda_a\tilde{\psi}_{\rm V}(\vec{p}_{\pm}) 
|N\rangle \, , \qquad {\rm for}\quad l=3,8\, .
\end{eqnarray*}
\begin{eqnarray*}
g_{1,\pm}^0 & = & N_C \frac{M_N}{\pi} 
\int_{M_N |x_{\mp}|}^{\infty}  p dp \sum_{\mu \ne{\rm V}}\Bigg\{
\\ & & \hspace{1em}
Q_{\mu}\left[\frac{\cos{\Theta^{\pm}}}{4}
\left(\sqrt{2} N_{\mu}^{(1)} + N_{\mu}^{(2)}\right)\mp
\frac{1}{8}\left(\sqrt{2} M_{\mu}^{(1)\pm}+M_{\mu}^{(2)\pm}\right)\right]
  \langle N |\Omega_{3} | N \rangle  
\\ & & \hspace{1em}
+ P_{\mu} \left[\frac{\cos\Theta^{\pm}}{2}
\left(\sqrt{2} N_{\mu}^{(1)}+N_{\mu}^{(2)}\right)\mp
\frac{1}{4}\left(\sqrt{2} M_{\mu}^{(1)\pm}+M_{\mu}^{(2)\pm}\right)\right]
 \langle N | D_{83} | N \rangle \Bigg\} \\ 
g_{1,\pm}^l & = & N_C \frac{M_N}{\pi} 
\int_{M_N |x_{\pm}|}^{\infty}  p dp\, \Bigg\{ \left[ 
  \tilde{g}_{\rm V}\tilde{f}_{\rm V}\cos{\Theta^{\pm}} \mp 
  \frac{1}{2} \tilde{g}_{\rm V}^2  \mp
  (\cos^2\Theta^\pm-\frac{1}{2}) \tilde{f}_{\rm V}^2\right] 
  \langle N | D_{l3} | N \rangle 
\\ & & \hspace{1em}
+ 2 \sum_{\mu \ne {\rm V}}  \Bigg( \delta_{G0} \Gamma_{\mu} 
\left[\frac{\cos\Theta^{\pm}}{2} 
\left(\tilde{g}_{\rm V} \tilde{f}_{\mu}^{(1,+)} +
\tilde{f}_{\rm V} \tilde{g}_{\mu}^{(1,+)} \right) \mp 
\frac{1}{2}\tilde{g}_{\rm V} \tilde{g}_{\mu}^{(1,+)} 
\mp(\cos^2\Theta^\pm-\frac{1}{2})\tilde{f}_{\rm V}\tilde{f}_{\mu}^{(1,+)}
\right]
\\ && \hspace{12em}\times
\langle N |  D_{l3}(D_{88}-1) |  N \rangle \\
& & \hspace{1em}
+ \frac{\delta_{G1}}{\sqrt{3}} 
  Q_{\mu} \left[ \frac{\cos\Theta^\pm}{4}
\left(\sqrt{2} N_{\mu}^{(1)} + N_{\mu}^{(2)}\right)
  \mp\frac{1}{8}\left(\sqrt{2}M_{\mu}^{(1)\pm}+M_{\mu}^{(2)\pm}\right)
\right] \langle N | D_{l8} \Omega_{3}| N \rangle 
\\ & & \hspace{1em}
+ \frac{\delta_{G1}}{\sqrt{3}} 
  P_{\mu}\left[\frac{\cos\Theta^\pm}{2}
  \left(\sqrt{2} N_{\mu}^{(1)}+N_{\mu}^{(2)}\right)
  \mp\frac{1}{4}\left(\sqrt{2}M_{\mu}^{(1)\pm}+M_{\mu}^{(2)\pm}\right)
\right]  \langle N| D_{l8} D_{83} | N \rangle 
\\ & & \hspace{1em}
+ \frac{1}{4} \frac{{\cal N}^2_{n0}}
{\epsilon_{\rm V}-\epsilon_{\mu}}\delta_{l_\mu 0}
\Bigg[\left(\tilde{g}_{\rm V} \cos\Theta^\pm \mp 
\frac{1}{2} \tilde{f}_{\rm V}(2\cos^2\Theta^\pm-1)\right)
\\ && \hspace{12em} \times
\left(\bar{w}_{n0}^+ \bar{w}_{n0}^- \tilde{g}_{\rm V}(k_{n0}) 
\tilde{t}_{n0}^1+\bar{w}_{n0}^{-2} \tilde{f}_{\rm V}(k_{n0})
\tilde{t}_{n0}^1\right)
\\ & & \hspace*{7em} +\frac{1}{2}
\left(\tilde{f}_{\rm V}\cos\Theta^\pm \mp\tilde{g}_{\rm V}\right)
\left(\bar{w}_{n0}^{+2} \tilde{g}_{\rm V}(k_{n0}) \tilde{s}_{n0} +
\bar{w}_{n0}^{+} \bar{w}_{n0}^{-} \tilde{f}_{\rm V}(k_{n0}) 
\tilde{s}_{n0}\right) \Bigg]
\\* && \hspace{12em} \times
\langle N|2d_{3\alpha\beta}D_{l\alpha}\Omega_{\beta}|N\rangle 
\\ & &  \hspace{1em}
+\frac{m-m_s}{2 \sqrt{3}}  
 \frac{{\cal N}^2_{n0}}{\epsilon_{\rm V}-\epsilon_{\mu}} 
\Bigg[\left(\tilde{g}_{\rm V}\cos\Theta^\pm \mp 
\frac{1}{2}\tilde{f}_{\rm V}(2\cos^2\Theta^\pm-1)\right)  
\\ && \hspace{12em} \times
\left(\bar{w}_{n0}^+ \bar{w}_{n0}^- \tilde{g}^{\Theta_c}_{\rm V}(k_{n0}) 
\tilde{t}_{n0}^1-\bar{w}_{n0}^{-2}\tilde{f}^{\Theta_c}_{\rm V}(k_{n0})
\tilde{t}_{n0}^1\right) \\
& &  \hspace*{7em} +\frac{1}{2}
\left(\tilde{f}_{\rm V}\cos\Theta^\pm \mp\tilde{g}_{\rm V}\right) 
\left(\bar{w}_{n0}^{+2} \tilde{g}^{\Theta_c}_{\rm V}(k_{n0}) 
\tilde{s}_{n0}-\bar{w}_{n0}^{+}\bar{w}_{n0}^{-} 
\tilde{f}^{\Theta_c}_{\rm V}(k_{n0}) \tilde{s}_{n0}\right)
\Bigg] 
\\ && \hspace{12em}\times
\langle N | 2 d_{3 \alpha \beta} D_{l \alpha} D_{8 \beta} | N \rangle 
\Bigg) \Bigg\} \\   
\end{eqnarray*}
for $l=3,8$. The summation involves  $i=1,2,3$ as well as
$\alpha,\beta=4,\ldots,7$. We have used the following shorthand 
notation in order to simplify the presentation
\begin{eqnarray*}
N^{(j)}_{\mu} & = & 
\tilde{g}_{\rm V} \tilde{f}^{(j,-)}_{\mu} + \tilde{f}_{\rm V}
\tilde{g}^{(j,-)}_{\mu} \, ,\qquad j=1,2 \\
M^{(1)\pm}_{\mu} & = &  
\left(\tilde{g}_{\rm V} \tilde{g}^{(1,-)}_{\mu} 
(3\cos^2\Theta^\pm-1)+\tilde{f}_{\rm V}\tilde{f}^{(1,-)}_{\mu}\right) 
\left(1+\cos^2\Theta^\pm\right) \\ 
M^{(2)\pm}_{\mu} & = & 2 
\left(\tilde{g}_{\rm V} \tilde{g}^{(2,-)}_{\mu} + \tilde{f}_{\rm V}
\tilde{f}^{(2,-)}_{\mu}(2\cos^2\Theta^\pm-1)\right) 
\end{eqnarray*}
while Q$_{\mu}$, $P_{\mu}$ and $\Gamma_{\mu}$ are again those
listed already in appendix B.

\medskip
\leftline{\large\it C.4 Explicit expressions for $\Delta$g$_1$}
\medskip
Finally we present the formulae for the bilocal correction
$\Delta g_1$.
$$
\Delta g_1 = \delta_{0} \Delta g_{1,+}^0 + \delta_{3} \Delta g_{1,+}^3 +
\delta_{8} \Delta g_{1,+}^8 + \delta_{0}' \Delta g_{1,-}^3 +
\delta_{3}' \Delta g_{1,-}^3 + \delta_{8}' \Delta g_{1,-}^8 \, .
$$
Herein the expression for the singlet is quite simple
\be
\Delta g_{1,\pm}^0 &=& \frac{N_C}{2 \pi} \frac{\partial}{\partial x}
    \int_{M_N |x_{\pm}|}^{\infty} p dp\,
 \left[\cos\Theta^\pm \tilde{g}_{\rm V}(p)
  \tilde{f}_{\rm V}(p) \mp \frac{1}{2} \tilde{g}_{\rm V}(p)^2 \mp
 (\cos^2\Theta^\pm-\frac{1}{2})\tilde{f}_{\rm V}(p)^2 \right]
\nonumber \\ && \hspace{4cm}\times
 \left[\frac{1}{2 \alpha^2} - \frac{\alpha_1}{\alpha^2} \langle N | D_{83} | N
   \rangle \right]
\nonumber
\ee
while for $l=3,8$ we find 
\begin{eqnarray}
\Delta g_{1,\pm}^3 &=& \frac{N_C}{2 \pi} \frac{\partial}{\partial x}
    \int_{M_N |x_{\pm}|}^{\infty} p dp\,
 \left[\cos\Theta^\pm \tilde{g}_{\rm V}(p)
  \tilde{f}_{\rm V}(p) \mp \frac{1}{2} \tilde{g}_{\rm V}(p)^2 \mp
 (\cos^2\Theta^\pm-\frac{1}{2})\tilde{f}_{\rm V}(p)^2 \right] 
\nonumber \\ &&\hspace{7em} \times
 \Bigg[-\frac{1}{\beta^2} \langle N | d_{3 \alpha \beta} D_{3\alpha} 
R_{\beta} | N \rangle - \left( \frac{\beta_1}{\sqrt{3} \beta^2} 
+\frac{1}{6 \beta^2} - \frac{1}{4 \alpha^2} \right) 
\langle N | D_{33} | N \rangle   
\nonumber \\ && \hspace*{8.5em} 
\nonumber \\ && \hspace*{8.5em} 
-\frac{1}{\sqrt{3}} \left( -\frac{\beta_1}{\beta^2}
+\frac{\alpha_1}{\alpha^2}\right) \langle N | D_{38} D_{83} | N \rangle 
+\frac{\beta_1}{\sqrt{3} \beta^2} \langle N | D_{33} D_{88} | N \rangle 
\nonumber \\ && \hspace*{8.5em}
-\frac{1}{\sqrt{3} \alpha^2}\langle N|D_{38}R_3|N\rangle\Bigg]\, ,
    \nonumber 
\end{eqnarray}
\vskip1cm
\begin{eqnarray}
\Delta g_{1,\pm}^8  &=& \frac{N_C}{2 \pi} \frac{\partial}{\partial x}
    \int_{M_N |x_{\pm}|}^{\infty} p dp\,
 \left[\cos\Theta^\pm \tilde{g}_{\rm V}(p)
  \tilde{f}_{\rm V}(p) \mp \frac{1}{2} \tilde{g}_{\rm V}(p)^2 \mp
 (\cos^2\Theta^\pm-\frac{1}{2})\tilde{f}_{\rm V}(p)^2 \right] 
\nonumber \\* &&\hspace{7em} \times
 \Bigg[-\frac{1}{\beta^2} \langle N | d_{3 \alpha \beta} D_{8\alpha} 
R_{\beta} | N \rangle - \left( \frac{- \beta_1}{\sqrt{3} \beta^2} 
+\frac{1}{6 \beta^2} - \frac{1}{4 \alpha^2} \right) 
\langle N | D_{83} | N \rangle   
\nonumber \\* && \hspace*{8.5em} 
\nonumber \\* && \hspace*{8.5em} 
-\frac{1}{\sqrt{3}} \left( -2 \frac{\beta_1}{\beta^2}
+\frac{\alpha_1}{\alpha^2}\right) \langle N | D_{88} D_{83} | N \rangle 
-\frac{1}{\sqrt{3} \alpha^2}\langle N|D_{88}R_3|N\rangle\Bigg]\, .
    \nonumber
\end{eqnarray}
This completes the list of matrix elements entering our 
structure function calculation.

\end{document}